\documentclass{egpubl}
\usepackage{vcbm2021}
 
\WsSubmission      % uncomment for submission to EG Workshop
\usepackage[T1]{fontenc}
\usepackage{dfadobe}  
\usepackage{amsmath}
\usepackage{amsfonts} 
\usepackage{mathptmx}
\usepackage{amssymb}
\usepackage[tight,normalsize,sf,SF]{subfigure}
\usepackage{todonotes}

\biberVersion
\BibtexOrBiblatex
\usepackage[style=alphabetic, bibstyle=EG, labelnumber, defernumbers=true,  backend=biber, backref=true]{biblatex}

\DeclareFieldFormat{labelnumberwidth}{\mkbibbrackets{#1}}
\renewbibmacro*{cite}{%
  \printtext[bibhyperref]{%
    \printfield{labelprefix}%
    \ifkeyword{secondary}
      {\printfield{labelnumber}}
      {\printfield{labelalpha}%
      \printfield{extraalpha}}}}

\defbibenvironment{bibliographyNUM}
  {\list
     {\printtext[labelnumberwidth]{%
        \printfield{labelprefix}%
        \printfield{labelnumber}}}
     {\setlength{\labelwidth}{\labelnumberwidth}%
      \setlength{\leftmargin}{\labelwidth}%
      \setlength{\labelsep}{\biblabelsep}%
      \addtolength{\leftmargin}{\labelsep}%
      \setlength{\itemsep}{\bibitemsep}%
      \setlength{\parsep}{\bibparsep}}%
      }
  {\endlist}
  {\item}
  
\electronicVersion
\PrintedOrElectronic
\usepackage{egweblnk} 
\title[Towards Narrative Medical Visualization]%
      {Towards Narrative Medical Visualization}

\author[M. Meuschke, L. Garrison, N. Smit, S. Bruckner, K. Lawonn, B. Preim]
{\parbox{\textwidth}{\centering M. Meuschke$^{1,2}$, L. Garrison, N. Smit, S. Bruckner, K. Lawonn$^{2}$, B. Preim$^{1}$}
        \\
{\parbox{\textwidth}{\centering $^1$Department of Simulation and Graphics, University of Magdeburg, Germany \\
    $^2$Institute of Computer Science, University of Jena, Germany\\
      }
}
}
\begin{document}

% uncomment for using teaser
% \teaser{
%  \includegraphics[width=\linewidth]{eg_new}
%  \centering
%   \caption{New EG Logo}
% \label{fig:teaser}
%}

\maketitle
%-------------------------------------------------------------------------
\begin{abstract}
%% ============================================================================
Narrative visualization aims to communicate scientific results to a general audience and garners significant attention in various applications. 
%% ============================================================================
Merging exploratory and explanatory visualization could effectively support non-expert
understanding of scientific processes. 
%% ============================================================================
Medical research results, e.g., mechanisms of the healthy human body, explanations of pathological processes, or avoidable risk factors for diseases, 
are also interesting to a general audience that includes patients and their relatives.
%% ============================================================================
% Medical knowledge is immediately relevant for patients and their relatives but 
% also other groups benefit from medical knowledge. 
%% ============================================================================
% Besides, medical student education may benefit from meaningful integration of 
% narrative techniques into representations of medical scenarios.
%% ============================================================================

%% ============================================================================
This paper discusses how narrative techniques can be applied to medical visualization to tell data-driven stories about diseases. 
%% ============================================================================
% Our targeted audiences comprise four classes: medical students, medical-related 
% people such as athletes, patients and their relatives, and broad public. 
% We describe medical scenarios 
We address the general public comprising people interested in medicine without specific medical background knowledge. We derived a general template for the narrative medical visualization of diseases.
%% ============================================================================
Applying this template to three diseases selected to span bone, vascular, and organ system, we discuss how narrative techniques can support visual communication and facilitate understanding of medical data. 
%% ============================================================================
Other scientists can adapt our proposed template to inform an audience on other diseases.
%Focused on data-driven visualizations, we discuss the pros and cons of narrative 
% visualizations for medical outreach.   
%% ============================================================================
With our work, we show the potential of narrative medical visualization and 
conclude with a comprehensive research agenda.


\end{abstract}  
%-------------------------------------------------------------------------
\section{Introduction}
\label{sec:intro}
%% ============================================================================
%% ============================================================================
% Traditionally, interactive data exploration (exploratory visualization)
% is a paradigm for supporting domain experts in their discovery and analysis processes,
% while explanatory visualization is a paradigm for broad audiences who are
% the recipients of the results emanating from experts’ exploration. 

% Since the arrival of the WorldWideWeb, interactive data exploration has increasingly 
% been available to broad audiences, empowering them to explore and analyze data independently. 

% The general availability of open data and powerful computers is thus enabling a confluence
% of exploratory and explanatory visualization, and in fact a cross-fertilization of the two. 

% In a recent viewpoint article, Ynnerman et al. [9] elaborate on this confluence
% and coin the term Exploranation to denote this new paradigm in science
% communication and its feedback impact on traditional exploratory paradigms, as it is
% noted that also exploratory visualization is increasingly making use of explanatory
% methodology.
% In this new landscape of technology and methodology for visual communication
% to broad audiences, there is a wide range of knowledge areas that need to be mastered
% to develop successful visualization beyond the traditional tools for domain experts.

%% ============================================================================
Medical visualization research to date has focused primarily on supporting medical experts (radiologists, pathologists, surgeons in diagnosis and treatment and---to a lesser extent---to medical students, in particularly for anatomy education. Medical information and research, however, are also interesting to non-experts, i.e., a general audience that comprises patients and their relatives along with those with an interest in science. Interactive medical visualization aiming at this type of audience requires different design approaches with easy to understand representations~\cite{Bottinger2020} than in systems such as radiology workstations that are aimed at experts. 
%To reach a general audience, %with little familiarity with the underlying 
%science or visualizations, motivating and 
%interesting representations need to be designed~\cite{Bottinger2020}. 
%% ============================================================================

Narrative visualization combines storytelling techniques with interactive graphics to appeal to a general audience~\cite{Segel2010}. It aims to present the data in a traceable progression that is memorable and easier to understand~\cite{Figueiras2014}. There are two types of storytelling: synchronous and asynchronous storytelling~\cite{Lee2015}. In synchronous storytelling, the narrator is in direct contact with the audience, e.g., live presentations, whereas asynchronous stories do not require direct audience contact. These stories take the form of recorded videos, static graphics, % are presented to the audience. 
%% ============================================================================
% While asynchronous storytelling includes both visually guided tours through 
% complex processes and interactive visualizations, these concepts are only 
% applicable to a small audience in synchronous scenarios~\cite{Ynnerman2020}.
%Asynchronous storytelling also includes 
or visually guided tours through complex processes with interactive visualizations. 
%% ============================================================================
% In case the size of the audience is rather small interactive visualizations can 
% also help to communicate data in synchronous scenarios~\cite{Ynnerman2020}. 
%% ============================================================================

%% ============================================================================
Ynnerman et al.~\cite{Ynnerman2018} coined the term \emph{exploranation} for merging exploratory visualizations that are traditionally made for experts with explanatory visualization techniques. While this supports visual knowledge acquisition for non-experts, it requires more guidance and automatically-generated content. For medical imaging data this may comprise labeled visualizations and animations highlighting relevant anatomical structures, e.g., vessel branching around an associated pathological structure. Visualizations of population-based data that indicate the most frequent tumor locations or metastasis pathways may also interesting to the public. Moreover, visualizations of health survey data demonstrating avoidable lifestyle-related risk factors for diseases can motivate the public to adopt healthier lifestyles.
%are potentially interesting for general audience seeking a healthier way of living. 
%% ============================================================================

%% ============================================================================
While \emph{“scientific outreach”} is already an essential topic for the visualization of astronomy data~\cite{Bock2019}, climate data~\cite{Bottinger2020reaching}, and cell biology data~\cite{Kouvril2021}, the same has not been true for interactive medical visualization research. Exceptions include epidemiological data, e.g., the COVID-19 Dashboard by Johns Hopkins University which supports map-based visualization, a selection of interesting countries, and time-based visualization of cases and fatalities. Early limited authoring tools were developed for generating interactive medical stories based on volume data~\cite{Wohlfart2006,Wohlfart2007}. However, medical data also includes other data types, e.g., clinical images, 3D models, and flow data. 
%% ============================================================================

%% ============================================================================
Several techniques for visualizing medical imaging data lend themselves well to narrative medical visualization storytelling principles with limited freedom for exploration. 
%Narrative medical visualization involves storytelling principles for generating visualizations with limited freedom for exploration. This may involve 
These include clipping planes which are automatically moved, cutaways or 
automated ghosted views based on structure selection, and automatically generated animated transitions. %, adapting the camera position 
%and orientation as well as visual styles. 
%% ============================================================================
% We aim at data-driven visualizations instead of purely artistically generated 
% visualizations. 
%% ============================================================================
%% ============================================================================
%The basis for these visuals are data-driven representations.
% We aim at a flexible and scaleable approach to generate data-driven 
% visualizations, making it possible that the visualization author can easily refine or adapt such a visualization. 
%% ============================================================================
However, concept-driven content, e.g., informative infographics, may be highly valuable to engage general 
audiences in scientific communication~\cite{Rheingans2020}.
%% ============================================================================

\begin{figure}[t!]
  \centering
  \includegraphics[width=\linewidth]{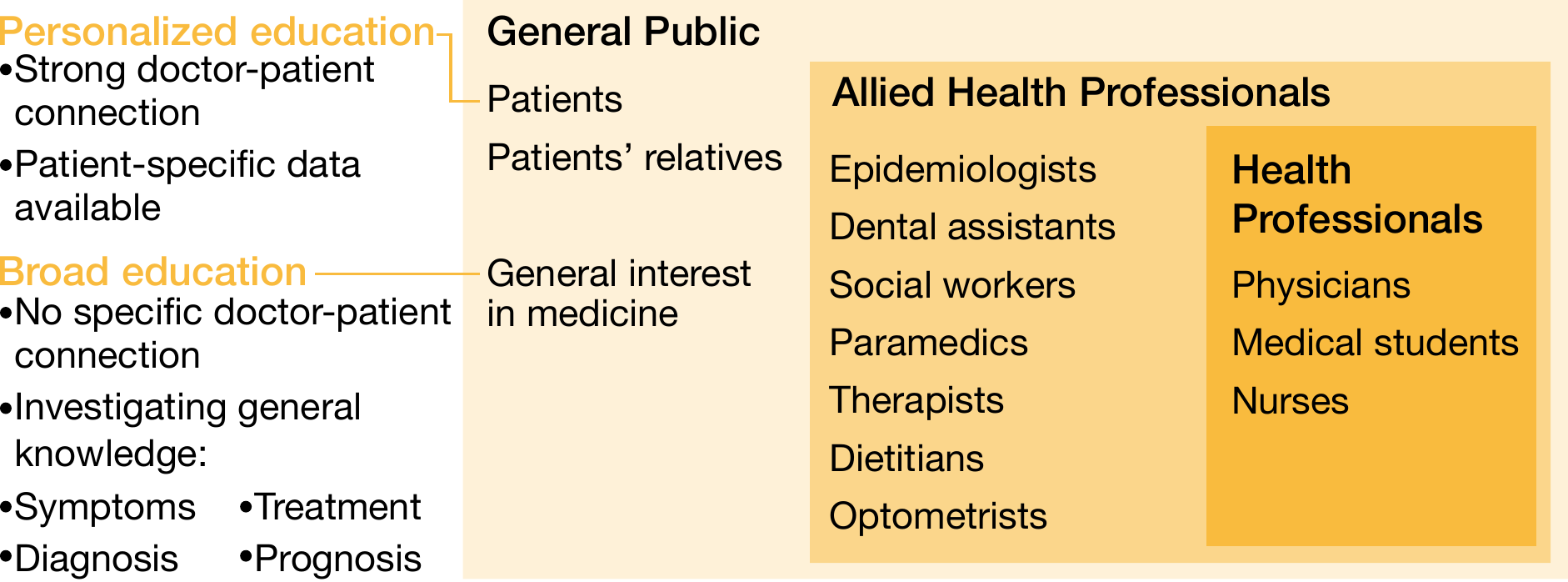}
  \caption{\label{fig:overview_audiences} Audiences affected by narrative medical visualization. }
  \vspace{-15px}
\end{figure}

%% ============================================================================
\noindent
\textbf{Scope of this Paper.} In this work, we discuss the potential of including interactive exploration of medical data in narrative visualization for a general audience, i.e., members of the general public who are interested in understanding diseases and their treatment but lack detailed medical knowledge or familiarity with scientific visualizations.
%% ============================================================================
%We assume that this audience will not have detailed medical knowledge or be familiar with scientific visualizations. 
%% ============================================================================
We further identify three general public subgroups: Patients with a direct
link to a specific disease, patient relatives, and people interested in 
medicine, %without a direct connection to the disease, 
see Figure~\ref{fig:overview_audiences}. 
%% ============================================================================
%In this paper, we focus on designing stories for medically interested people.
%% ============================================================================
%% ============================================================================
%Using an asynchronous storytelling method, we discuss how narrative techniques can be used to present medical data in an understandable way. 
Following an asynchronous storytelling method, we show how to leverage narrative techniques to present medical data in a way that is both compelling and understandable. Our proof-of-concept focuses on the suitability and arrangement of narrative techniques to tell stories about 
%% ============================================================================
%However, we focus on which techniques are suitable and how they should be arranged instead of what an efficient authoring tool should look like. 
%% ============================================================================
%With a focus on the visualization of diseases, we select 
three common diseases that are related to three important structures of the human body: organs, vessels, and bones. Our inspiration for these disease stories draws in part from health websites such as \textit{WebMD} and \textit{UpToDate}. Similar to other works dealing with narrative scientific visualization, we choose touch screen as a medium such that the user can interact with the data during the story. 
In narrative visualization, stories can be mainly data-driven or concept-driven. We follow the suggestions by Segel and Heer~\cite{Segel2010} that data should enrich the story while memorable visuals and interesting storytelling are the main components of the story.
%% ============================================================================
Our key contributions are the following: 
\begin{itemize}
    \item We provide an overview of existing work in narrative visualization and based on an analysis of a corpus of 30 medical stories propose a template to structure medical visualization stories.
    \item We present three proof-of-concept medical stories that are enriched with interactive medical data visualization components to explain information around selected example diseases.
    \item We identify promising areas for future research in narrative medical visualization.
\end{itemize}
\noindent
\textbf{Organization.} Section~\ref{sec:ingrnarrvis} summarizes 
general narrative techniques based on seminal works in this field. 
%% ============================================================================
Then, Section~\ref{sec:narrscivis} gives a brief insight which other scientific 
visualization areas have used narrative techniques and how. 
%% ============================================================================
Here, we also describe the associated transition from scientific
visualization designed for experts to scientific visualizations for 
the general public, as well as challenges which arise in this process.  
%% ============================================================================
Section~\ref{sec:exampmedstory} then describes the core of our paper. 
%% ============================================================================
Based on the summary in Section~\ref{sec:ingrnarrvis}, 
we show how narrative techniques can be applied to medical data to generate stories for the general public.
%% ============================================================================
We then discuss various aspects of our conceptualized medical stories in 
Section~\ref{sec:discussion} and identify a research agenda that highlights promising aspects for future work in medical narrative visualization in Section~\ref{sec:agenda}. 
%% ============================================================================
The paper is concluded in Section~\ref{sec:conclusion}.

\section{Ingredients of Narrative Visualization}
\label{sec:ingrnarrvis}
%% ============================================================================
A visual data story is composed of a series of specific facts, called 
\emph{story pieces}, that are supported by data~\cite{Lee2015}. 
%% ============================================================================
These story pieces are visualized to convey important messages to the audience. 
%% ============================================================================
Visualizations are enriched with story elements such as labels, arrows, 
links, and textual explanations to clearly emphasize these messages and avoid 
ambiguity.
%% ============================================================================
The story pieces should be arranged into scenes on the basis of a meaningful 
genre and design pattern 
%transitions should be inserted between them 
to support the author's communication goal. 
%% ============================================================================
General goals include to inform or entertain the audience. %, but may also include the creation of orientation, motivation and feedback.
%% ============================================================================
In the following, we summarize existing techniques to generate and transition between scenes. 
%% ============================================================================
In addition, we summarize genres and design patterns with suggestions for their use in medical visualization.  

\subsection{Generating and Transitioning Narrative Scenes}
%% ============================================================================
Segel and Heer~\cite{Segel2010} 
%analyze 58 representations 
derive general 
design elements of narrative visualizations and examine the range of user 
guidance and interaction. 
%% ============================================================================
% They identified seven genres of narrative visualization: magazine style, 
% annotated chart, partitioned poster, flow chart, comic strip, slide show, and 
% film/video/animation, and presented three design patterns for interactive stories: 
% the martini glass structure, interactive slideshows, and drill-down stories. 
% %% ============================================================================
% With this they made a first steps towards a better understanding of visual data 
% stories.
%% ============================================================================
Hullman and Diakopoulos~\cite{Hullman2011} build on this work to
analyze 51 narrative visualizations, examining the rhetorical devices used. % in narrative visualizations. 
%% ============================================================================
Stolper et al.~\cite{Stolper2016} extend this summary by novel data-driven 
storytelling techniques. % that have emerged since 
%these articles were published. 
%% ============================================================================
% Based on an analysis of 45 web-based visualizations, they organized existing 
% narrative techniques into four high-level categories: communicating narrative
% and explaining data, linking separated story elements, enhancing structure and navigation, and providing controlled exploration.
%% ============================================================================
Based on these ground-breaking works in the field of narrative visualization, 
we summarize existing story elements, how to connect them to form scenes, how to 
transition between scenes and how to construct a scene path.
%% ============================================================================

%% ============================================================================
\subsubsection{Story Elements}
%% ============================================================================
%For storytelling, v
Visualizations are best complemented by other means of communication and highlighting techniques need to guide the user through a story~\cite{Kosara2013}. 
%% ============================================================================
%The authors has different ways to communicate the story and to explain data. 
%% ============================================================================

\noindent
\textbf{Text narration.} Text is the simplest way to explain data.  
%% ============================================================================
%Text can be used in different versions. 
%% ============================================================================
\textit{Long-form texts} can be used to explain key points in 
detail and to introduce or summarize a topic. 
%% ============================================================================
\textit{Headlines} or \textit{captions} can serve to draw attention to a story. 
%% ============================================================================
\textit{Tooltips} can provide details when 
a user hovers their cursor over an element~\cite{Figueiras2014}.
%% ============================================================================
Text can also be used in the form of \textit{annotations} %in visualizations such 
%as diagrams or in the form of
or \textit{labels} to designate important structures. 

\noindent
%% ============================================================================
\textbf{Audio narration} can be used to enhance visualizations~\cite{Segel2010}. 
%% ============================================================================
This allows the viewer to focus more on the visuals, since the narrative is 
temporally linked to the visual elements. 
%% ============================================================================
% However, the use of audio makes it more difficult for the viewer to regulate the 
% pace of the story. 
%% ============================================================================

\noindent
Moreover, \textbf{graphical properties} can be used to draw the reader’s attention. 
%% ============================================================================
Elements can be highlighted using wrapped shapes, specific colors or techniques 
such as motion or close ups~\cite{Segel2010}. 
%% ============================================================================
% With these approaches key observations can be pointed out which is an important 
% aspect in narrative visualization.  

\subsubsection{Connection of Story Elements}
%% ============================================================================
% Telling a data-driven story usually requires several story elements such as 
% texts, charts, maps or 3D images. 
%% ============================================================================
To understand the explanatory nature of the interplay between story elements, 
connections must be made between them. 
%% ============================================================================
Stolper et al.~\cite{Stolper2016} found three basic types to connect story elements.
%% ============================================================================

\begin{figure*}[t!]
  \centering
  \includegraphics[width=\linewidth]{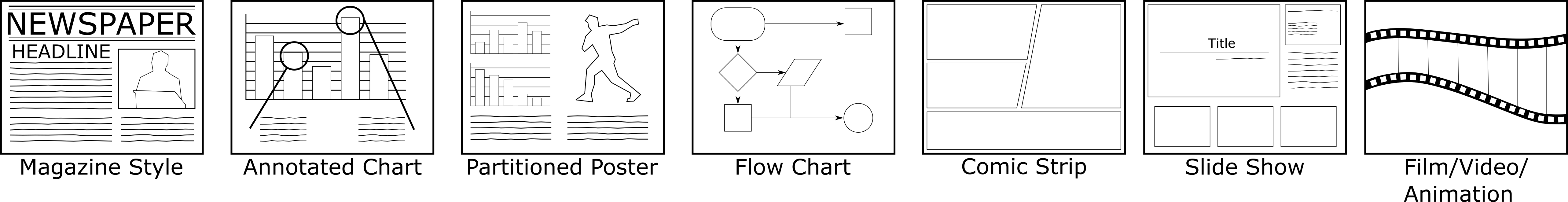}
  \caption{\label{fig:genre} General genres of narrative visualization according 
  to Segel and Heer~\cite{Segel2010}.} %, Copyright IEEE. }
  \vspace{-10px}
\end{figure*}

\noindent
%% ============================================================================
\textbf{Interaction} is an efficient way to connect story elements. 
%% ============================================================================
Interactivity refers to the different ways a user can manipulate the 
visualization, e.g., filtering, hovering, zooming, rotating, and translating  %selecting, searching, 
%navigating), 
and also to how the user learns these methods (explicit instruction, 
tacit tutorial, initial configuration)~\cite{Segel2010,Stolper2016}.
%% ============================================================================
% While these techniques can also be applied to 3D representations, in 3D space 
% objects can also be rotated or translated along the third axis. 
%In 3D space, objects can also be rotated or moved along the third axis. 
%% ============================================================================

%% ============================================================================
%The degree of interaction can be divided into different modes. 
%% ============================================================================
% Wohlfart and Hauser~\cite{Wohlfart2007} distinguished four modes of how strongly 
% a user can interact with the visualization. 
%% ============================================================================
The level of interaction ranges from passive narration, where no interaction 
is provided, to free exploration, where the user has no interaction 
constraints~\cite{Wohlfart2007}. 
%% ============================================================================
%In between, two other modes are distinguished. 
%% ============================================================================
Passive narration can be interrupted and the user can temporarily take control 
and change the presentation, e.g., by using \textbf{dynamic queries} to change the 
visual style of an object. 
%% ============================================================================
Afterwards, the passive narration continues. 

\noindent
%% ============================================================================
\textbf{Color} is another option to link story elements. 
%% ============================================================================
Consistent colors should be used to represent objects or attributes that 
appear in multiple visualizations~\cite{Stolper2016}.
%% ============================================================================
Color can also be used to connect text and visualizations by assigning text 
the same color as the associated visualized objects. 
%% ============================================================================
However, the choice of color schemes and the design of the color map play a 
crucial role. 
%% ============================================================================
Crameri et al.~\cite{Crameri2020} presented guidelines to design color charts 
for scientific data, including perception effects to be considered. 
%w.r.t. the quantities actually visualized, the audience, and the message 
%to be communicated.

% \cite{Bottinger2020}

% Although originally intended for the meteorological domain, the set of guidelines for
% effective colormaps presented by  are mostly generally applicable.
% Furthermore, potential psychological effects of the colormap design used need to
% be taken into account especially with respect to the actual quantities visualized,
% the target group and the narrative to be communicated. Schneider and Nocke~\cite{Schneider2018}
% explored the impact of using different color schemes for the same temperature change
% visualization. On the basis of a user study, they found the original blue–red–magenta
% color scheme to be more alerting than the other schemes analyzed; but at the same
% time, it caused disillusioning associations of powerlessness and fear—feelings that
% are undesired if engagement of the audience is a desired communication goal.

\noindent
\textbf{Animations} can also be used to link objects that help users relate 
complex processes in an understandable way. 
%% ============================================================================
Care must be taken to ensure that the user does not lose the focus while 
context information is needed for orientation. 
%% ============================================================================
Therefore, smooth transitions between different camera positions are required, 
where focus objects should be visually emphasized. % and context objects should be 

\subsubsection{Defining Scene Transition}
%% ============================================================================
Moving within and between visual scenes without disorienting the user is a 
fundamental aspect of storytelling. 
%% ============================================================================
Segel and Heer~\cite{Segel2010} identified six types of transitions. %, which we 
%summarize below.   
%% ============================================================================
One way %to maintain the user's orientation between scenes 
is to keep the object change between scenes to a minimum, maintaining \textit{object continuity}. 
%% ============================================================================
The number and style of objects should not be fundamentally changed between 
two cuts. 
%% ============================================================================
Related to this is the concept of \textit{familiar objects}, which states that 
commonly used symbols should be used to represent facts.
%% ============================================================================
Another category involves meaningful movement of the virtual camera. 
%% ============================================================================
The \textit{view angle} of the camera should change between two scenes or when 
moving within a scene, but not so much that completely different views are created. 
%% ============================================================================
Also, strong changes in the \textit{camera movement speed} between adjacent 
scenes should be avoided. 
%% ============================================================================
% However, the speed can increase or decrease continuously over several scenes to 
% generate a dynamic.
%% ============================================================================
\textit{Continuity editing} is an established technique from the film industry 
which creates the impression that the story was shot in one piece without cuts. 
%% ============================================================================
%This requires the use of multiple cameras and special camera effects. 
%% ============================================================================
% \textit{Object continuity} refers to the representation of objects. 
% %% ============================================================================
% Colors and other visualization features should be kept constant for the same 
% objects throughout the story. 
%% ============================================================================
Another option is to use \textit{animated transitions}. 
%% ============================================================================
Based on morphological transformations, objects of one scene can be changed 
into objects of another scene. % via various intermediate stages. 
%% ============================================================================

%% ============================================================================
%% ============================================================================
\subsubsection{Defining a Scene Path}
%% ============================================================================
%% ============================================================================
% Compared to traditional data exploration, where the user is not given a path in 
% which order to look at the data, 
Data-driven stories are usually characterized by an author-specified order. 
%% ============================================================================
The story is thus given a structure that is supported by frequent navigation aids. 
%% ============================================================================
In addition to the specification of a strict path (linear story), there is the 
possibility to provide the user with several paths to choose from (user-directed story)~\cite{Segel2010}. 
%% ============================================================================

%% ============================================================================
Commonly used techniques %to allow the viewer 
to navigate through a story are 
\textit{next/previous buttons} and \textit{scrolling}. 
%% ============================================================================
\textit{Flowchart arrows} can help to convey the intended narrative 
structure of the story. % to the user.
%% ============================================================================
To navigate to a specific location, \textit{menu selections} or 
\textit{interactive maps} %, e.g., for geographic data, 
can be provided. % in the story. 
%% ============================================================================
To show the user where s/he is in the story \textit{section header buttons}, 
\textit{breadcrumbs} in the form of points, and \textit{timelines} in the form 
of progress bars or checklists are often used. 
%% ============================================================================
% Similarly, \textit{timelines} in the form of progress bars or checklists indicate
% how much content the user has already seen from the story. 
%% ============================================================================
%% ============================================================================

%% ============================================================================
%% ============================================================================
\subsection{Selecting Narrative Genres}
%% ============================================================================
%% ============================================================================
To communicate the story in an understandable way, consideration must be given 
to how story elements are arranged and combined.  
%% ============================================================================
Segel and Heer~\cite{Segel2010} have defined seven genres: 
magazine style, annotated chart, partitioned poster, flow chart, comic strip, 
slide show, and film/video/animation, as depicted in Figure~\ref{fig:genre}. 
%% ============================================================================
These genres differ in the number of scenes shown, and the arrangement of story 
elements within a scene.
%% ============================================================================
%Not every genre is equally well suited for storytelling. 
%% ============================================================================
The choice of genre depends on the data complexity, as well as the intended 
audience and medium. 
%% ============================================================================

%% ============================================================================
For narrative medical visualization, \textit{magazine styles}, where a 2D image 
is embedded in text, could be adapted to integrate 3D models, with the text 
around explaining visible structures.
%% ============================================================================
% Instead of using a static 2D image, interactive 3D models could be used, where 
% the text around it could explain details about visible structures. %with which the user can interact (rotation, translation, zooming).  
%% ============================================================================
% Depending on the currently viewed body region and displayed structures (organs, vessels, bones, muscles, cellular structures) within the zoom level, the text around it could explain details about the visible structures. 
%% ============================================================================
In contrast, \textit{flow charts} can be used to show medical processes such as 
disease treatment in an abstract way. 
%% ============================================================================
% However, the individual states could be made clickable to provide 
% the user with more detailed information in the form of magazines. 
%% ============================================================================
%% ============================================================================
\textit{Annotated charts} can be used to present statistical information, e.g., 
the %distribution of a disease in different cohorts or 
prognosis as a function of the selected therapy. 
%% ============================================================================
The combination of images and diagrams in a \textit{partitioned poster} is well 
suited to provide overviews or summaries of medical explanations. 
%% ============================================================================
% They can draw the reader's attention to a topic by highlighting interesting 
% aspects such as the increasing incidence of a disease. 
%% ============================================================================
% Similarly, advantages and disadvantages of treatment strategies can be summarized comparatively.
%% ============================================================================
%% ============================================================================
\textit{Slide shows} are commonly used in business presentations. %, where a presenter
%explains the content of each slide to the audience. 
%% ============================================================================
For the application to medical data, the user's attention should be kept by 
interactive components, where the s/he is encouraged to interactively explore the data. % to gain additional information. 
% where the user is supposed to explore the data interactively, it would be helpful to replace the presenter with an audio guide that explains the content of the slides. 
% %% ============================================================================
% To keep the user's attention during the slide show, the audio guide could actively encourage the user to interact with the interactive components of the slides to gain additional information. 
%% ============================================================================
%% ============================================================================
\textit{Comic strips} consist of highly abstracted illustrations that contain 
only brief annotations. %and tend to have a humorous background. 
%% ============================================================================
An interesting scenario would be the cartoon-style illustration of medical aspects for children. % as target audience. %Conceivable, for example, would be the depiction of first aid measures. 
%% ============================================================================
%% ============================================================================
\textit{Videos, and animations} would be well suited to support the 
exploration of 3D medical data. %, as the perception of information in this case is highly dependent on the views chosen. 
%% ============================================================================
Optimal views on surfaces, such as vessels and organs, could show structures of interest, e.g., 
the resection of a tumor. % in a comprehensible way. 
%% ============================================================================
%Animations could also show disease development over time. % such as the bulging of a vessel (aneurysm) or the narrowing of a vessel (stenosis) which can lead to a stroke. 
%% ============================================================================
% To actively involve the user in the exploration, he should be able to pause videos/animations to interact with the objects himself. 

%\vspace{-5px}
%% ============================================================================
%% ============================================================================
\subsection{Selecting Narrative Design Patterns}
\label{subsec:narpatterns}
%% ============================================================================
%% ============================================================================
Depending on author intent and the audience, a story can be told in 
different ways. 
%% ============================================================================
Bach et al.~\cite{Bach2018} described eighteen narrative design patterns that 
can be used individually or in combination to tell data stories. % in a variety of ways. 
%% ============================================================================
Each pattern has a specific purpose, with five overarching groups. 
%% ============================================================================
There are \textit{argumentative}, \textit{structuring}, \textit{framing}, 
\textit{emotional}, and \textit{engaging} patterns. 
%% ============================================================================
% However, a pattern can also belong to several groups. 
% %% ============================================================================
% Below, we briefly summarize the design patterns and provide insights into how 
% they could be used for medical storytelling. % based on medical data.  
%% ============================================================================

%% ============================================================================
%\noindent
Argumentative patterns include comparisons, concretizations, and repetitions, 
to present, support, reinforce, contradict, or discuss a particular statement.
%% ============================================================================
They can be used, for example, to compare treatment options, to present information 
that users should remember (e.g., preventable risk factors), or to 
present the benefits of protective measures, such as vaccinations.   
%% ============================================================================
Structuring patterns include concepts such as revealing, slowing down, and speeding up. 
%% ============================================================================
Framing patterns determine how the story content is perceived through 
techniques such as creating familiar settings, making guesses, defamiliarization,
breaking conventions, hiding data, and using physical metaphors. 
%% ============================================================================
Structuring and framing patterns are important to present medical data, 
which usually contains different data types, such as volume data, 3D 
models, quantitative values and qualitative flow information. %, which together describe complex processes in the body. 
%% ============================================================================
To communicate this data to general audiences, it must be simplified, details 
must be omitted, and interesting aspects, e.g., in statistical diagrams, 
should be revealed. 
%% ============================================================================
Slowing down, and speeding up could, e.g., be used to show blood flow 
animations. 
%% ============================================================================
In time ranges where interesting flow occurs, the animation is slowed 
down and in less interesting ranges the animation is sped up~\cite{Kolesar2014}. 
%% ============================================================================
%% ============================================================================
Emotional patterns such as directly assessing the audience and presenting individual 
stories are designed to help understand and share important feelings in the story.
%% ============================================================================
To engage the user into the story, techniques such as rhetorical questioning, 
call to action, and interactive exploration can be used.

\section{Narrative Visualization of Scientific Data}
\label{sec:narrscivis}
%% ============================================================================
Numerous works combined narrative techniques and information visualization~\cite{Tong2018,Gershon2001}. 
%% ============================================================================
In contrast, there is little research on combining scientific visualization 
with narratives~\cite{Ma2011}. 
%% ============================================================================
In this section, we summarize the transition from expert-driven visualizations of 
scientific data, i.e., spatio-temporal data, to non-expert visual representation of these data. 
%% ============================================================================
We also provide insights into the challenges that arise during this transfer, 
especially for medical data. 
%% ============================================================================
Finally, we present selected scientific applications outside of medicine, where narrative 
techniques have already been used. 
%% ============================================================================

%% ============================================================================
\subsection{From Scientific to General Audience}
\label{subsec:sciens_to_pub}
%% ============================================================================
%% ============================================================================
% The general goal of visualizations is to bring abstract or complex data into a 
% visually comprehensible form by omitting negligible details. 
%% ============================================================================
Traditionally, visualizations were used by experts to gain detailed insights into 
complex data. 
%% ============================================================================
Experts have a deep background knowledge of the respective domain and are able 
to evaluate and interact with complex visualizations. 
%% ============================================================================
% They want to understand underlying processes and test hypotheses 
% in order to generate knowledge. 
%% ============================================================================
 % with some practice. 
%% ============================================================================
% They use visualization to validate data, where the results and insights are shared 
% with other experts. % using visualization techniques in scientific publications and through presentations at professional conferences. 
%% ============================================================================

%% ============================================================================
In contrast, a general audience includes people with varying levels of expertise who %from the general public. 
%% ============================================================================
%These people that are assumed to 
%have a broad range of knowledge and %education, as 
%well as to 
differ in terms of age and cultural backgrounds~\cite{Bottinger2020}. 
%% ============================================================================
Bringing scientific results to a general audience is challenging, 
as it [...] “is quite a different matter to compel attention and understanding
in a diverse, hurried, skeptical population of readers than to communicate with an
eager, familiar group of associates”~\cite{Dibiase1990}.
%% ============================================================================
Therefore, the purpose of the visualization should be clearly defined in the 
context of the target audience in order to fulfill the intended communication goals~\cite{Bottinger2020}.
%% ============================================================================

%% ============================================================================
% In order to communicate results to broad audiences requires the craft of visual 
% storytelling with data. 
Results from cognitive science show that embedding data in a narrative makes it 
more exciting and memorable~\cite{Ma2011}. 
%% ============================================================================
%In order to arouse the interest of a broad public, 
For this purpose, complex scientific results need to be reduced, summarized and 
generalized by means of simplified and understandable visualizations. % in order to arouse the 
%interest of a broad public. 
%% ============================================================================
Compromises have to be made in terms of accuracy and completeness, since showing too many details %, such as visualizing several quantities simultaneously or 
%annotate too many structures, 
can make it difficult to convey a clear message~\cite{Bottinger2020}. 
%% ============================================================================
%% ============================================================================

%% ============================================================================
% The reason for this is the existence of two types of memory: semantic memory, to 
% remember individual facts, and episodic memory, to remember sequences of events. 
% %% ============================================================================
% By enriching visualizations with narrative techniques, episodic memory can be activated. 
% %% ============================================================================
% As a result, the representations are memorized as a coherent sequence~\cite{Ma2011}.
%% ============================================================================
% However, visualizations can appear biased by omitting details. 
% %% ============================================================================
% To gain confidence in the representations, information about the data sources and 
% which parts of the data are visualized should be provided~\cite{Bottinger2020}. 

%% ============================================================================
To create a narrative visualization, the target audience must be defined as 
precisely as possible~\cite{Bottinger2020}. 
%% ============================================================================
The background knowledge and the goals of the audience are decisive for the design, 
the level of interaction allowed and how strongly the audience is guided through the 
story. 
%% ============================================================================
With regard to medical data, different audience groups such as scientists, students, 
patients, health care providers, or policymakers are conceivable, as shown in Figure~\ref{fig:overview_audiences}. 
%% ============================================================================

\subsection{Challenges in Narrative Visualization}
%% ============================================================================
%% ============================================================================
Several challenges need to be considered when designing narrative visualizations for rich scientific data~\cite{Bock2019,Ynnerman2020}. 
%% ============================================================================
We summarize the main challenges and their relation to medical data.
%% ============================================================================

% We are still in the early days of exploranation in public spaces, and much research
% needs to be conducted to support this area, which has the potential to become one
% of the largest uses of visualization with impact reaching far beyond the traditional
% research and development domains targeted by visualization research~\cite{Ynnerman2020}.

% In developing visualization for public spaces, it is imperative to closely connect
% with state-of-the-art research in science education. It is also apparent that learning
% research has largely in the past been dealing with cognitive and sociocultural aspects
% of visual representation and multimedia learning [11, 19, 23]. Only recently, studies
% gathering and evaluating digital platforms, games, simulations, animations and
% traditional visualizations in teaching and learning have been conducted [7, 8, 22].
% From this literature, we learn that the content itself, how it is presented, people’s
% prior knowledge, interest and engagement, also affect the result~\cite{Ynnerman2020}.

\noindent
\textbf{Varying Spatial-Temporal Scales.} In scientific data, the spatial and 
temporal scales of objects can vary greatly~\cite{Bock2019}.  
%% ============================================================================
Navigation and interaction aids are needed that identify points of interest both 
spatially and temporally. %, and display multiple representations of the same object 
%depending on the chosen scale. 
%% ============================================================================
In medical data, the sizes of structures can vary greatly. % can be represented as hierarchical structures, with the sizes of 
%objects changing at each hierarchy level.  
%% ============================================================================
For example, organs, such as the liver, are several centimeters in diameter, 
while embedded structures, such as vessels or cells, are many times smaller. 
%% ============================================================================
Similarly, time scales can range from hours that a treatment needs to years in 
long-term follow-up of diseases. 

% For example, organs such as the liver  can be represented as a whole organ at the top level.
%% ============================================================================
%At the top level is the outer tissue envelope of the liver. 
% At lower levels, smaller smaller structures such as vessels are encoded, which 
% contain smaller cell-based components that make up the blood. 
%% ============================================================================
% Within the liver are smaller structures such as vessels, which also contain 
% smaller components that make up the blood. 
%% ============================================================================

%% ============================================================================
\noindent
\textbf{Varying Data Sources.}
%% ============================================================================
Another problem are different types of data coming from different 
sources~\cite{Bock2019}. 
%% ============================================================================
Medical data can include radiological and histological image data, numerical 
values, and statistical information, which can be acquired with different devices, 
e.g., different scanners. 
%% ============================================================================
Moreover, biomedical simulation data may be relevant. 
%including derived representations such as flow trajectories, which 
%that characterize the blood flow behavior. % over the cardiac cycle.
%% ============================================================================

%% ============================================================================
\noindent
\textbf{Data Access Issues.} Another challenge that is particularly relevant to 
medical data is making data available to the general public. 
%% ============================================================================
From an ethical point of view, mere anonymization of data is not sufficient 
for their use in public scenarios. 
%% ============================================================================
One solution to this could be to use data derived from data donors. 
%% ============================================================================

%% ============================================================================
\noindent
\textbf{Interaction and Navigation.}  
%% ============================================================================
The exploration of medical data by the general public requires to reduce 
complexity in terms of interaction and navigation compared to  
systems for experts~\cite{Ynnerman2020}. 
% BP: An expert system is about artificial intelligence
%% ============================================================================
Otherwise, users can lose their desire to use the visualizations. 
%% ============================================================================
The design of the user interface should be tailored to the communication goal of 
the story without noticeably restricting the exploration. 
%% ============================================================================

\noindent
\textbf{Occlusion management.} In 3D scenes, special attention must be paid to 
resolve object occlusion. 
%% ============================================================================
% Elmqvist and Tsigas~\cite{Elmqvist2008} identified five design patterns to manage
% occlusion within 3D scenes: multiple viewports, virtual X-ray tools, tour planners, 
% volumetric probes, and projection distorters. 
%% ============================================================================
Virtual X-ray approaches and volumetric probes to adapt the opacity 
of occluding objects either automatically or interactively~\cite{Elmqvist2008} are 
suitable for narrative medical visualization.
%% ============================================================================
Whenever interesting objects are occluded in medical data, often smart visibility 
techniques, such as ghosted views and cut away techniques, are 
applied~\cite{lawonn2018}. 
%% ============================================================================
% Moreover, Elmqvist and Tsigas classify visual tasks affected by occlusion 
% comprising discovering, assessing and spatial relating of targets as well as 
% creating, deleting, and modifying objects. 
% %% ============================================================================
% For narrative medical applications, the tasks of discovering and assessing targets 
% are very important.  
%% ============================================================================

%% ============================================================================
\noindent
\textbf{Storytelling and Exploration.} To not overwhelm people with visual exploration opportunities, they should be guided through the story~\cite{Ynnerman2020}. 
%% ============================================================================
%Meaningful interaction and navigation techniques are needed for this. 
%% ============================================================================
In 3D medical visualization, this can be realized through automatic views, 
limited rotation capabilities, or predefined parameter settings. 
%% ============================================================================
The user should always know where s/he is in 3D space. %, which requires the display 
%of focus and context information. 
%% ============================================================================

%% ============================================================================
\noindent
\textbf{Flexibility and Performance.} Due to different data types and many 
possible scenarios, %that can be presented by different devises, 
a system to interactively explore medical data should be flexible regarding the 
integration of new interactions and rendering styles. 
%% ============================================================================
In addition, robustness is important to make software 
available to the general public. 

\subsection{Selected Examples of Narrative Scientific Visualization }
%narrative techniques~\cite{Ma2011}. %to make interesting aspects 
%of the data accessible to the general public. 
%% ============================================================================
%% ============================================================================
Typical places where the general public comes into contact with 
scientific data are museums, planetariums, exhibitions, and science centers. 
%% ============================================================================
Ma et al.~\cite{Ma2011} described projects of NASA's Scientific Visualization 
Studio in which narrative visualizations communicate investigations recorded 
with various instruments and sensors. % to the general public. 
%% ============================================================================
Further details are provided by captions, sound or live demonstrations. 
%% ============================================================================
Media comprise UltraHD displays and hyperwalls, Dome shows, mobile, 
and 360 projections. 
%% ============================================================================
% Storytelling enables the user to interact with geographic data such as the Earth’s climate or the collapse of a star by using a story model, such as story nodes or story transitions~\cite{Akiba2009}. 
%% ============================================================================

%% ============================================================================
Krone et al.~\cite{Krone2017} present design considerations of a scientific 
exhibition in the Carl-Zeiss-Planetarium Stuttgart to inform 
a general audience about computer simulations comprising industrial and molecular 
simulation examples. 
%% ============================================================================
In different interactive scenarious, the users are educated what simulation is, 
how they are computed and how the results can be visualized. 
%% ============================================================================
A Microsoft Kinect and Leap Motion are used as input device. 
For validation, the visitors could provide feedback using questionnaires about 
different aspects of the exhibition. 
%% ============================================================================
Although only a few visitors left feedback, this was very positive with regard 
to the comprehensibility and engagement of the presentations shown. 
%% ============================================================================

%% ============================================================================
Recently, Ynnerman et al.~\cite{Ynnerman2020} summarized how storytelling is 
used in the Norrk\"oping Visualization Center C. 
%% ============================================================================
In addition to dome projections and VR setups, users can explore volume data 
using multi-touch displays. 
%% ============================================================================
These data comprise full-body CT scans, which are visualized by direct volume 
rendering (DVR). 
%% ============================================================================
% First, an effective algorithm was developed to represent the memory-intensive 
% volume data~\cite{Ljung2006}. 
% %% ============================================================================
% Subsequently, this method was combined with an easy-to-use multi-touch interface. 
%% ============================================================================
From a pre-defined image gallery, the visitors can select a transfer 
function, which should be applied to the data. % without complicated parameter adaption. 
%% ============================================================================
Visitors can interact directly with the visualizations, they can perform single 
and multi touch gestures, e.g, rotate the volume or cut through it, but they cannot select objects very precisely.
%The visitor can also rotate the volume or cut through it using a clipping plane. 
%% ============================================================================
Similar setups are used to explore a virtual human mummy~\cite{Ynnerman2016} and 
biological structures that would not be visible to the naked eye~\cite{Host2018}.
%% ============================================================================
However, besides pre-defined transfer functions, textual descriptions, and videos no further guidance through the complex data is provided. % to the user. 
Narrative techniques have also been used to communicate potential future climate 
changes to the general public based on simulated data~\cite{Bottinger2020}.
%% ============================================================================
In order for the user to draw conclusions, %from the changes shown, 
various visualization aspects must be taken into account. 
%% ============================================================================
The choice of appropriate color scales is important to draw the user's 
attention. 
%% ============================================================================
Furthermore, combinations of visualization techniques, such as color and contour 
lines to show correlations, must be carefully explained, e.g., by audio guidance. 

\section{Narrative Medical Visualization Concepts}
\label{sec:exampmedstory}
%% ============================================================================
%% ============================================================================
In this section, we describe how narrative techniques can enrich medical 
visualization so that users are able to easily understand, absorb, and interact with the data. 
%% ============================================================================
Our intended \textbf{communication goal} is to \textbf{inform} 
\textbf{people interested in medicine} about a disease. 
%% ============================================================================
% In terms of users, we focus on \textbf{people interested in medicine} with no 
% specific medical background knowledge, 
% cf. Section~\ref{subsec:sciens_to_pub}. 
%% ============================================================================
%These comprise two main groups: \textbf{patients} and \textbf{people interested in medicine}. 
%% ============================================================================
To demonstrate the potential of narrative medical visualization, we first derive 
a template comprising potential stages of a story about disease data, as detailed in Section~\ref{subsec:stagesmedstory}. 
%% ============================================================================
Next, in Section~\ref{subsec:medusecases}, we define an example scenario where the target audience comes into contact 
with medical data.
% Afterwards, we define an example scenario, where medically interested people , see 
% Section~\ref{subsec:medusecases}.
%% ============================================================================
Then, we select three common diseases for story generation, discussed in  Section~\ref{subsec:meddata}, 
followed by explaining the story preparation including data preprocessing and 
selection of an authoring tool in Section~\ref{subsec:medicalstorydesign}.
%% ============================================================================
Based on the defined template and selected diseases, finally the medical 
stories are designed and presented in Section~\ref{subsec:narmedvis_broad}.

\begin{figure*}[t!]
  \centering
  \includegraphics[width=\linewidth]{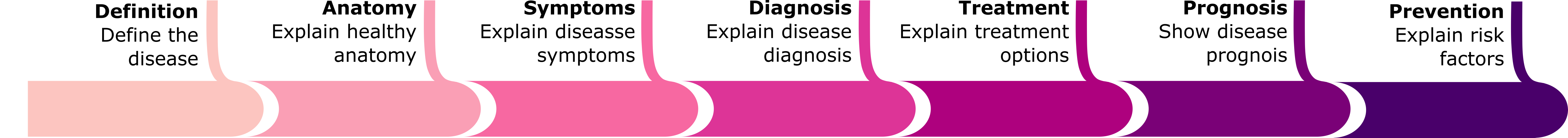}
  \caption{\label{fig:stages} Derived template for narrative medical visualization 
  of disease data that comprises seven stages.}
  \vspace{-10px}
\end{figure*}

%% ============================================================================
\subsection{A Template for Narrative Disease Visualization}
\label{subsec:stagesmedstory}
%% ============================================================================
Many university hospitals, scientific institutes or online encyclopedias %such as 
%Wikipedia or Radiopaedia 
put freely accessible blogs online to inform a general 
audience %without specific medical background knowledge 
about the development, diagnosis and treatment of various diseases. 
%% ============================================================================
% On the Internet you can find numerous reviews and videos that provide information 
% about diseases you are looking for.  
% %% ============================================================================
% The aim of such explanatory blogs is to provide an overview of various disease 
% aspects such as diagnosis and treatment options. 
%% ============================================================================
% The target audience here is people who do not have specific medical background 
% knowledge. 
%% ============================================================================
We analyzed a total of 30 blogs for three selected diseases: Liver cancer \cite{surgery_liver_metas_1}-\nocite{surgery_liver_metas_2, surgery_liver_metas_3, surgery_liver_metas_4,surgery_liver_metas_5,surgery_liver_metas_6,surgery_liver_metas_7,surgery_liver_metas_8,surgery_liver_metas_9}\cite{surgery_liver_metas_10}, 
% Liver cancer(\cite{surgery_liver_metas_1,surgery_liver_metas_2, surgery_liver_metas_3, surgery_liver_metas_4, surgery_liver_metas_5,surgery_liver_metas_6,surgery_liver_metas_7,surgery_liver_metas_8,surgery_liver_metas_9,surgery_liver_metas_10})
brain aneurysm \cite{aneurysm_1}\nocite{aneurysm_2,aneurysm_3,aneurysm_4,aneurysm_5,aneurysm_6,aneurysm_7,aneurysm_8,aneurysm_9}-\cite{aneurysm_10},  
% (\cite{aneurysm_1,aneurysm_2,aneurysm_3,aneurysm_4,aneurysm_5,aneurysm_6,aneurysm_7,aneurysm_8}), 
and pelvic fracture
\cite{pelvic_1}\nocite{pelvic_2,pelvic_3,pelvic_4,pelvic_5,pelvic_6,pelvic_7,pelvic_8,pelvic_9}-\cite{pelvic_10} 
% (\cite{pelvic_1,pelvic_2,pelvic_3,pelvic_4,pelvic_5,pelvic_6,pelvic_7,pelvic_8,pelvic_9,pelvic_10}) 
according to their 
basic structure, since we have the same communication goal and want to address 
the same audience. 
%for our 
%similarly-structured narrative presentations. 
%% ============================================================================

%% ============================================================================
The basic structure of these blogs is very similar. 
%% ============================================================================
First, a short and understandable \textbf{definition of the disease} is provided, 
and \textbf{statistical aspects} such as the annual incidence and age-related 
distribution between men and women are described. 
%% ============================================================================
Next, an \textbf{anatomical overview} shows the location and function 
of the structures affected by the disease. 
%% ============================================================================
This provides the baseline for understanding what is normal before introducing 
the disease itself. 
%% ============================================================================
Using the example of liver cancer, schematic sketches are used to explain 
where the liver is located, its function, and important nearby structures. 
%% ============================================================================
% Subsequently, the known \textbf{causes of the disease} are explained. 
% %% ============================================================================
% This is often coupled to schematic drawings that show in a highly simplified way 
% how the disease ultimately develops.  
%% ============================================================================

%% ============================================================================
Subsequently, typical \textbf{symptoms} are explained usually as textual
enumeration.
%% ============================================================================
Afterwards, the \textbf{diagnosis} is explained. 
%% ============================================================================
This comprises frequently used examination methods, e.g., MRI, as well as their sequence and reliability in order to make a diagnosis. 
%% ============================================================================
The procedure of each diagnostic method is briefly summarized and associated 
inconveniences for the patient are explained. 
%% ============================================================================

%% ============================================================================
The diagnosis is typically followed by an overview of possible 
\textbf{treatment options}. 
%% ============================================================================
Therapeutic procedures are summarized including treatment risks and the 
associated chance of cure is estimated. 
%% ============================================================================
Typically, 5-year \textbf{prognoses} are provided. 
%% ============================================================================

%% ============================================================================
Finally, \textbf{disease prevention} is explained, where risk factors for the 
development are summarized. 
%% ============================================================================
A distinction is usually made between \emph{preventable} and \emph{congenital/genetic} risk 
factors. 
%% ============================================================================
This concluding consideration of risk factors serves as an appeal and clarification that one's own behavioral patterns can have a strong influence on the development of life-threatening disease. 
%% ============================================================================
The reader should be sensitized to think about their own habits and to adapt 
their lifestyle in a positive way. 
%% ============================================================================
% Especially for certain cancers, this section also refers to screening 
% examinations and their importance in early detection. 
%% ============================================================================

%% ============================================================================
Based on this analysis, we derived a sequence of seven stages forming a template, as shown in Figure~\ref{fig:stages}, 
that can be used as a basic pattern for applying narrative techniques 
to disease data.
%% ============================================================================
% \begin{enumerate}
% \item Disease definition %combined with statistical aspects regarding incidence and 
% %age-distribution
% \item Anatomy %Anatomical overview of affected structures 
% \item Symptoms
% \item Diagnosis
% \item Treatment
% \item Prognosis
% \item Prevention
% \end{enumerate}
%% ============================================================================
% This sequence can be used as a basic pattern for applying narrative techniques 
% to disease data to inform people interested in medicine. 
%% ============================================================================
%- \textbf{\color{red}Uebersicht der analysierten Blogs und deren Struktur}
%% ============================================================================
% - search for examples in newspaper (New York Times) --> ask Laura 
% - analyze youtube videos for liver cancer and heart disease 

%% ============================================================================
%\subsection{Exemplary Selected Medical Scenarios}
\subsection{General Public Information}
\label{subsec:medusecases}
People interested in medicine come typically into contact with medical data 
either in museums or science centers or through Internet research on their home computers, where no specific doctor patient connection exists. 
%% ============================================================================
Similar to the Visualization Center in Norrköping~\cite{Ynnerman2020}, asynchronous storytelling based on a touch display could be used to interactively inform about diseases. %in terms of their incidence, diagnosis, therapy as well as 
%preventable risk factors. 
%% ============================================================================
While users at home would probably be more likely to use a tablet or their phone, larger interactive displays could be used in a science center or museum.  
%% ============================================================================
% Narrative techniques can help bring data about diseases, consisting of medical 
% image data, statistical information, and quantitative values, to a wide audience in an understandable way. 
%% ============================================================================
%% ============================================================================
% We show how narrative techniques can support auto didactic learning at home or 
% in a science center, where people interested in medicine want to learn about 
% diseases.
We show how narrative techniques can help medically interested people to inform themselves about diseases at home or in a science center.
%% ============================================================================
This is an asynchronous scenario where the user interactively explores the data on their own.  
%% ============================================================================
% In contrast to the scenario of personalized education, anatomical variations or 
% degrees of severity of a disease hardly play a role. 
%% =========================================================================
The goal is to give a general overview of a disease, not focusing on anatomical variations or severity of a disease.
%% =========================================================================
% Therefore, a database with several data sets per disease is not needed for the 
% broad education, but only one exemplary data set per disease.
%% =========================================================================

%% ============================================================================
\subsection{Selected Disease Examples}
\label{subsec:meddata}
%% ============================================================================
% We want to show how narrative visualization can support the interactive exploration 
% of medical data to understand anatomical structures and related diseases. 
%% ============================================================================
For selecting example diseases, we oriented ourselves to the basic structures 
of the human body, which are visible in radiological image data: 
organs, vessels, and bones. Below, we outline our motivation for selecting three specific disease examples.
%% ============================================================================

\noindent
\textbf{Liver Cancer.} Regarding organ diseases, we selected liver cancer. 
%% ============================================================================
%While this disease was relatively rare, 
The number of new cases 
of liver cancer has doubled in the last 35 years~\cite{Liu2019}.  
%% ============================================================================
Accordingly, the interest in learning more about this 
disease on the part of the general public %, including affected patients, 
is likely increasing. 
%% ============================================================================
%% ============================================================================
In Germany, approximately 8790 people (6160 men, 2630 women) are newly diagnosed with liver cancer each year. 
%% ============================================================================
The average age of onset is 69.9 years for men and 72.1 years for women. 
%% ============================================================================
The increase in annual new cases %observed in recent years 
is associated with an increasing number of patients with liver cirrhosis, the 
high rate of new hepatitis B infections, and increasingly frequent obesity. %, and type 2 diabetes~\cite{Moraga2017}. % mellitus. 
%% ============================================================================
%Men are two to three times more likely than women to develop liver cancer. 
%% ============================================================================

%% ============================================================================
\noindent
\textbf{Brain Aneurysms.} For vessel diseases, we selected brain aneurysms, 
which are localized dilations of the brain vessels. %, approximately 
%5\,mm to 3\,cm in diameter. 
%% ============================================================================
In addition to older people, younger people are also frequently affected, making this condition of interest to the general public. 
%% ============================================================================
%% ============================================================================
%  Cerebral aneurysms are localized dilatations of the 
% blood vessels of the brain, approximately 5\,mm to 3\,cm in diameter. 
%% ============================================================================
About 3-5\,\% of all people likely have a brain aneurysm~\cite{Boulouis2017}. 
%% ============================================================================
In most cases, these aneurysms are found by chance and remain asymptomatic. 
%% ============================================================================
Brain aneurysms are more common in people over the age of 40, where women are 
affected more often than men by a ratio of 5:3.
%% ============================================================================
% It is interesting to note that the prevalence of aneurysms in adolescents is 
% vanishingly low and increases markedly only after the age of about 20 years. 
% %% ============================================================================
% This suggests that aneurysms are not congenital but acquired vascular malformations 
% in the course of life.
%% ============================================================================
Their greatest danger is that the vessel wall ruptures, which leads to internal bleeding. 
%% ============================================================================
The low incidence of rupture %, approximately 
%8000 cases per year, 
suggests that 80\,\% to 85\,\% of all brain aneurysms will never rupture~\cite{Singh2009}.
%% ============================================================================
Due to the low rupture rate and the existing treatment risks for the patient, 
physicians must limit treatment to high-risk patients.

\noindent
\textbf{Pelvic Fracture.} For bone-related diseases, we used a data set showing a pelvic fracture, which is a break in any of the bones that form the ring of 
bones at hip-level~\cite{Gordon2018}. 
%% ============================================================================
% Such a fracture can have many different causes and definitely affects young 
% patients as well as older persons. 
%% ============================================================================
% In the following, we go into more detail about the individual diseases and provide 
% an overview of their frequency, diagnosis, therapy, prognosis, and prevention. 
Severe cases show multiple fractures and/or an unstable fracture.
%% ============================================================================
This can occur as a result of high-energy trauma, e.g., car accident (20-22\,\%), 
or in frail or older patients from minor trauma, such as a fall (5-30\,\%). 
%% ============================================================================
High-energy trauma-related pelvic fractures often come with accompanying injuries that require immediate treatment.
%% ============================================================================
Pelvic fractures represent 3\,\% of skeletal injuries, with 5-16\,\% mortality 
(unstable pelvic fractures ~8\,\%).
%% ============================================================================
All age groups can be affected, but trauma has a higher incidence in young males, while in older populations is more associated with women.

\subsection{Medical Story Preparation}
\label{subsec:medicalstorydesign}
%% ============================================================================
For medical story design, some data preprocessing was necessary, which we detail in Section~\ref{subsubsec:dataprepro}. 
%% ============================================================================
Essentially, this involves preparation of radiological image data. % for scene generation. 
%% ============================================================================
We also had to choose an authoring tool to create scenes and define 
transitions between them, as discussed in Section~\ref{subsubsec:authoringtool}. 
%% ============================================================================

%% ============================================================================
\subsubsection{Data Preprocessing}
\label{subsubsec:dataprepro}
%% ============================================================================
For each story, an anonymous patient data set was used, comprising different 
radiological data types, such as ultrasound, MRI, and CT images. 
%% ============================================================================
As this data was anonymized, we had no access to patient-related meta information. % in form of a patient report. 
%% ============================================================================
Therefore, any introductory patient information at the beginning of 
each story is fictitious. 
%% ============================================================================
In the following, we shortly describe our data sets and necessary preprocessing steps. 
%% ============================================================================

%% ============================================================================
\noindent
\textbf{Liver Cancer.} 
%% ============================================================================
We used a data set of a patient with a stage 1 liver carcinoma provided 
by our clinical partners as sample data. 
%% ============================================================================
%The patient had severe abdominal pain and swelling of the liver could be palpated. 
%% ============================================================================
In this data, small and medium-sized tumors without lymph node involvement and metastases were diagnosed in the liver based on ultrasound and CT Angiography (CTA). 
%% ============================================================================
%No other tumors were found outside the liver. 
%% ============================================================================
In addition, the liver showed no cirrhotic changes. 
%% ============================================================================
% The patient is a heavy smoker but reported a normal alcohol consumption pattern 
% and is otherwise in good general health. 
%% ============================================================================
Due to multiple tumors, surgical removal was not an option. 
%% ============================================================================
Instead, the tumors should be treated with ablation.
%% ============================================================================
%% ============================================================================
Based on the CTA data, the liver as well as the tumors were segmented using \textit{HepaVision}~\cite{Bourquain2002}. 
%% ============================================================================
Moreover, other surrounding structures such as the heart and ribs were segmented and transformed to 3D surfaces. % meshes using Marching Cubes. 
%% ============================================================================

\noindent
\textbf{Brain Aneurysms.}
%% ============================================================================
For brain aneurysms, we used a data set from Berg et al.~\cite{Berg2015}, 
where a brain aneurysm was incidentally found during CTA. 
%% ============================================================================
The vasculature was segmented using the pipeline presented by M\"onch 
et al.~\cite{Moench2011a} and converted into a volume grid. %, which employs a threshold-based segmentation, followed 
%by a connected component analysis, and Marching Cubes to extract a 3D surface.
%% ============================================================================
% Then, a 3D surface is extracted via Marching Cubes applied to the binary 
% segmentation result and segmentation errors are manually corrected. 
%%============================================================================
Computational fluid dynamics (CFD) simulations were used to calculate the 
blood flow behavior. 
%%============================================================================
% For this purpose, the quality of the 3D surface is optimized~\cite{S97}, before 
% it is converted into a volume grid. 
%%============================================================================
Finally, particles are traced in the resulting vector field to depict the blood flow. 
%%============================================================================
All simulation details can be found in the work by Berg et 
al.~\cite{Berg2015}. 
%%============================================================================

%% ============================================================================
\noindent
\textbf{Pelvic Fracture.} We obtained a CT dataset of a woman with a pelvic 
fracture including pre-operative and post-operative scans. 
%% ============================================================================
We performed direct volume rendering (DVR) in 3D Slicer~\cite{Kikinis2014} in order to generate videos and images to support our story. 
%% ============================================================================
In addition, we provide interactive 3D scenes based on the Virtual Surgical Pelvis (VSP) model~\cite{smit2016pelvis}, which is in use as a virtual educational tool to teach pelvic anatomy~\cite{smit2016online}. 
%% ============================================================================
The VSP consists of anatomical surface models based on expert segmentation of a cryosection data set. 
%% ============================================================================
Selected VSP structures were embedded as interactive 3D models. 
%% ============================================================================

%% ============================================================================
\subsubsection{Authoring Tool Selection}
\label{subsubsec:authoringtool}
%% ============================================================================
We created the stories using \textit{PowerPoint 365 MSO} version 2105. 
%% ============================================================================
PowerPoint offers numerous possibilities to combine and visually arrange narrative elements. 
%% ============================================================================
Animations and transitions can be defined and different file formats, such as 
images, videos and interactive 3D models, can be integrated. 
%% ============================================================================
%Users also have the option to interact with the loaded 3D models during the story. 
%% ============================================================================
PowerPoint thus offers all the functionalities we need to design basic concepts that show the potential of narrative visualizations for medical data. 
%% ============================================================================
In addition, the wide availability of PowerPoint makes it a good choice for an interactive narrative visualization intended for a general audience.
%% ============================================================================

%% ============================================================================
\subsection{Medical Story Design}
\label{subsec:narmedvis_broad}
%% ============================================================================
We use the derived stages shown in Figure~\ref{fig:stages} as the basic 
structure for designing our medical stories.
%% ===========================================================================
Thus, some scenes for the three diseases are very similar. 
%% ===========================================================================
To avoid redundancy, we use the example of liver cancer to show a complete design of a medical story. % and interactions. 
%% ===========================================================================
For the two remaining diseases, we %created two storyboards that 
focus on illustrating disease-specific aspects. % and thus differ from the liver scenario. 
%% ===========================================================================
Visual placeholders are inserted for parts of the stories whose design is 
similar to the liver scenario. 
%% ===========================================================================
We attached all created stories as supplemental material.

\begin{figure*}[t!]
  \centering
  \includegraphics[width=\linewidth]{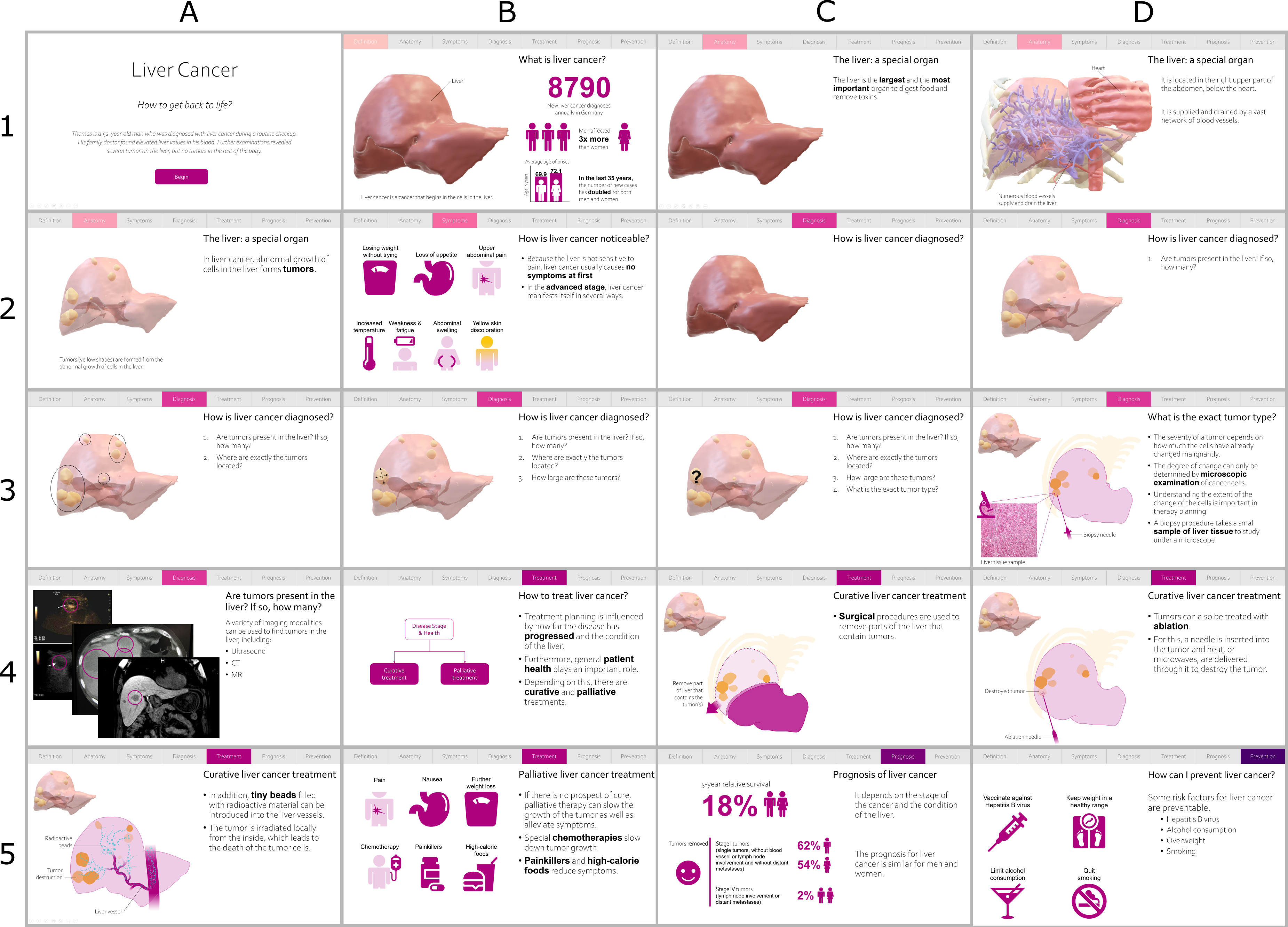}
  \caption{\label{fig:liver} All scenes of the liver cancer story 
  covering the seven derived stages. The narrative sequence is A1, B1, ..., D1, A2 and so on. }
\end{figure*}

%% ===========================================================================
We have chosen the \textit{slideshow} format as basic design genre. 
%% ============================================================================
%% ============================================================================
% The common thread of the story %in both scenarios 
% is based on the seven defined stages, see Figure~\ref{fig:stages}.  
%% ============================================================================
For each stage, one or multiple slides are prepared as scenes. 
%% ============================================================================
Within the scenes, other narrative genres such as magazine style or partitioned 
poster are used.
%% ============================================================================
Smooth transitions are defined between scenes, with the timeline of stages visible in each scene. 
%% ============================================================================
This way the user always knows s/he is in the story. 
All three stories introduce a patient case to capture the user's attention. 
%% ============================================================================
This consists of a catchy \textit{headline} and a \textit{long-form textual description} 
of the patient case, see Figure~\ref{fig:liver} (A1). 
%% ============================================================================
In addition, the patient description is read aloud as a \textit{voiceover}. 
%% ============================================================================
This patient description should help the audience to relate to the case and 
motivate them to continue with the story. 
%% ============================================================================
By pressing the start button, the user begins the actual story. 
%% ============================================================================
Below, we provide detailed insights into the scenes. 
%% ============================================================================

%% ============================================================================
\noindent
\textbf{Disease Definition.} Within the first scene, we use the 
\textit{magazine style} to introduce the affected anatomy by an interactive 3D 
model to provide an initial orientation to the topic. 
%% ============================================================================
Inspired by the work of Garcia~\cite{Arcia2016}, statistical parameters of the disease are depicted as \textit{information graphics} 
with \textit{icons} and laddered text to quickly absorb information. % without 
%becoming overwhelmed. 
%% ============================================================================
%This establishes the context of the disease. 
%% ============================================================================
We exclude visual representations of the disease at this stage to avoid 
overwhelming the user.
%% ============================================================================
%Liver cancer is a cancer that begins in the cells in the liver. 
%% ============================================================================
Via voice narration, the user is encouraged to rotate the model using their 
fingers. %to see all the tumors. 
%% ============================================================================
Since free rotation of 3D objects is difficult to handle for inexperienced users, 
we limit rotation around a vertical axis with a finger movement from left to right.
%% ============================================================================

%% ============================================================================
In the liver cancer story, we embed an interactive \textit{3D liver model} alongside 
\textit{textual components} and \textit{information graphics}, see Figure~\ref{fig:liver}~(B1). 
%% ============================================================================
We do not yet show 3D models of the tumors or unnecessary surrounding anatomy. 
%% ============================================================================
% A combination of labels and a caption is used to give a brief definition of liver 
% cancer. 
%% ============================================================================
%Statistical information is displayed next to the 3D model. 
%% ============================================================================
One of the most important statistical parameters is the annual incidence, which 
is emphasized by the larger font size. 
%% ============================================================================
Other parameters, such as the distribution between men and women and their 
average age of onset, are represented by \textit{annotated information graphics} 
and \textit{icons} to aid recognition and memorability.
%% ============================================================================
We apply similar concepts to the other two stories. 
%% ============================================================================
% Since the cerebral vasculature is very complex, the vascular model is embedded 
% in a semi-transparent representation of the skull bone to provide contextual 
% information. 
% %% ============================================================================
% Moreover, the aneurysm is highlighted by an enclosing circle and an arrow as it 
% otherwise could be overlooked in the vasculature.  
%% ============================================================================
%% ============================================================================

%% ============================================================================
\noindent
\textbf{Anatomy.} Again, we use the \textit{magazine style} to describe the anatomy scenes. 
%% ============================================================================
We describe the anatomical structures that are necessary to understand disease development. 
%% ============================================================================
These facts include the location, importance, and function of the key anatomical 
structure(s). 
%% ============================================================================
Finally, we introduce the disease by super- or juxtaposing the pathology on the 
normal anatomy with a crossfade transition. 
%% ============================================================================

%% ============================================================================
Continuing the liver cancer story, the anatomical stage defines four 
key facts: (1) the liver is the largest and the most important organ to digest 
food and remove toxins, (2) it is located in the right upper part of the abdomen, 
below the heart, (3) it is supplied and drained by a vast network of blood vessels, 
and (4) in liver cancer, abnormal growth of cells in the liver forms tumors. 
%% ============================================================================
We use several scenes to represent these key facts.
%% ============================================================================
% For the representation of the anatomy and important functions 
% of the affected structures we used several scenes. 

%% ============================================================================
% For each story the three to four most important characteristics of the 
% corresponding anatomy are described. 
% %% ============================================================================
% These facts contain also information about how the disease develops. 
% %% ============================================================================
% Each characteristic is represented by a scene. 
%% ============================================================================
First, an \textit{automatic rotation} of the 3D liver model gives an 
overview of its anatomical shape, where a \textit{long-form textual description} 
with \textit{highlighted keywords} provides more details. 
%In the first scene of this stage, an interactive 3D model of the healthy anatomy 
% is shown. 
%% ============================================================================
% Thus, a 3D model of the liver, the brain vasculature and the pelvis is presented, 
% respectively. 
%% ============================================================================
Following \textit{familiar objects} and \textit{object continuity} concepts, 
the story transitions to show anatomical context around the liver: 
\textit{labeled 3D models} of the ribs, heart, and liver vasculature. 
%% ============================================================================
%This provides anatomical context. 
%% ============================================================================
%To each structure a label is assigned. 
%% ============================================================================
% For the aneurysm case, this scene is skipped as there are no important 
% surrounding structures that are important for understanding the story. 
The last scene transitions to show the disease: surrounding structures 
\textit{fade away} and the fully-opaque liver becomes \textit{translucent} to 
reveal tumors within.
%% ============================================================================
The accompanying text discusses development of a liver tumor due to abnormal 
cell growth. 
%% ============================================================================

\begin{figure*}[t!]
  \centering
  \includegraphics[width=\linewidth]{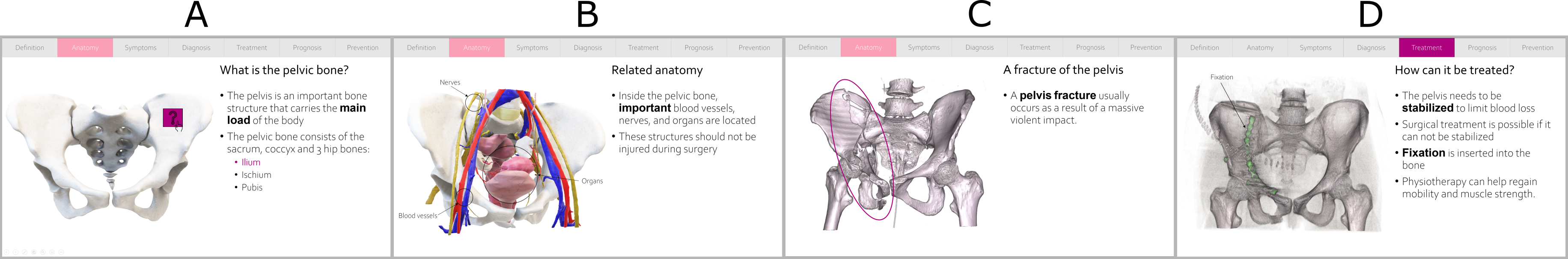}
  \caption{\label{fig:pelvis}Excerpts from the pelvic fracture story. A unique characteristic is the complex anatomical structure of the pelvis with multiple bones and closely related vessels, organs, and nerves. Hotspots, 3D models, and DVR are combined to highlight these aspects and treatment.}
   \vspace{-10px}
\end{figure*}

%% ============================================================================
The aneurysm and pelvic fracture stories both follow a similar introduction 
and flow of elements. 
%% ============================================================================
However, a unique characteristic is the complex anatomical structure of the 
pelvis with multiple bones and closely related vessels, organs, and nerves.
In addition, for this patient we have pre- and post-operative data available, which makes it possible to show treatment effects on real data.
%% ============================================================================
To communicate these anatomical peculiarities and the treatment process, we combine \textit{hotspots}, \textit{3D models}, and \textit{DVR}, see Figure~\ref{fig:pelvis}. 
%% ============================================================================
The user can interactively explore anatomical structures by clicking on the 
hotspots (A) by highlighting the corresponding anatomical name in the text 
description or clicking on a structure of interest. % on the right. 
%% ============================================================================
%With this anatomical structures can be interactively explored. 

% labeled substructures of 
% the focus structure, such as mayor supplying vessels in the brain or bone 
% structures of the pelvis, are blended in. 
%% ============================================================================
% For this purpose, artistic-driven techniques, such as dashed lines in the liver 
% case, are used to visually highlight the separation into substructures.
% the 3D surface models are divided into colored parts, with the 
% labels colored in the same way to create a visual connection. 
%% ============================================================================
% Besides, \textit{long-form textual description} explain the labeled structures 
% in more detail, where key words are highlighted in bold. 
%% ============================================================================

%at this point, the origin of an aneurysm is shown, 
%% ============================================================================
\noindent
\textbf{Symptoms.} Using the \textit{partitioned poster} genre, a visual overview of frequently-occurring symptoms is provided. 
%% ============================================================================
% To provide a visual overview about these symptoms we use the \textit{partitioned poster} 
% genre. 
%% ============================================================================
For each symptom, we artistically create an \textit{icon} with an accompanying 
\textit{caption}. 
%% ============================================================================
The use of icons as opposed to purely text-based listings aims to increase 
memorability of symptoms.% for the viewer. 
%% ============================================================================
The symptoms are displayed one after the other to give the user time to 
process each icon. 
%% ============================================================================

%% ================
% With regard to the symptoms, we have focused on frequently 
% occurring symptoms of the respective disease in order not to overwhelm the user 
% with too many exceptions. 
%% ============================================================================
For liver cancer, we create icons and accompanying text for the following 
critical symptoms in advanced liver cancer: unexplained weight loss, loss of appetite, 
pain/pressure in the upper abdomen, increased temperature, weakness/fatigue, 
abdominal swelling, and yellowing of the skin, see Figure~\ref{fig:liver} (B2). 
%% ============================================================================
We also identify key symptoms of pelvic fracture, with the same storytelling 
mechanisms. % we described above. 
%% ============================================================================

\begin{figure*}[t!]
  \centering
  \includegraphics[width=\linewidth]{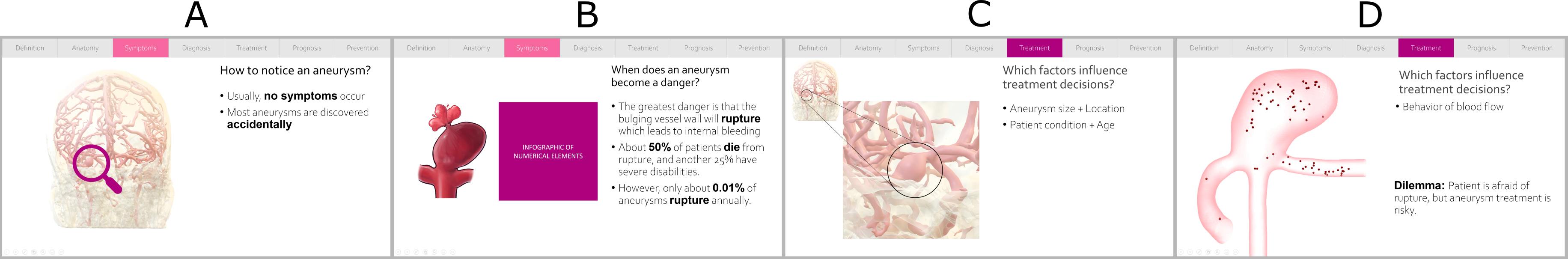}
  \caption{\label{fig:aneurysm}Excerpts from the aneurysm storyboard. There is a trade-off between the rupture risk and the treatment risk. 3D models, concept-driven visualizations, and blood flow depictions are combined to communicate this dilemma.}
    \vspace{-10px}
\end{figure*}

%% ============================================================================
We had to adjust the aneurysm story, since we focus on accidentally 
detected aneurysms without symptoms. %, we could not design symbols for symptoms. 
%% ============================================================================
%A unique characteristic of these aneurysms is that 
% Their detection represents a severe dilemma---it has to be communicated to the 
% patient and leads to fear of rupture. %is glorifying for the patient. 
%% ============================================================================
Treatment bears considerable risks, which can exceed natural rupture risk. % in everyday life.  
%% ============================================================================
Therefore, the aneurysm story communicates how rupture-prone aneurysms can be 
detected as shown in Figure~\ref{fig:aneurysm}. 
%% ============================================================================
%% ============================================================================
The first scene shows a 3D aneurysm model representing an incidental finding (see Figure~\ref{fig:aneurysm} (A)). 
%% ============================================================================
An illustrative superimposed \textit{magnifying glass} helps to quickly see the 
aneurysm in the complex vascular tree.  
%% ============================================================================
Next, aneurysm rupture is shown \textit{illustratively} (see Figure~\ref{fig:aneurysm} 
(B). 
%% ============================================================================
Using \textit{information graphics} and \textit{textual descriptions}, 
arranged in \textit{magazine style}, it should be made clear that a rupture occurs 
very rarely but is very dangerous. 
%% ============================================================================
Here, the information graphics are only indicated by a placeholder, as these 
would be similar to the graphics %for displaying statistical quantities as 
of the liver definition stage. 
%% ============================================================================

%% ============================================================================
\noindent
\textbf{Diagnosis.} The diagnosis is discussed using a variety of media. 
%% ============================================================================
Key to this stage is informing the audience on how the diagnosis is achieved 
comprising diagnostic questions and imaging modalities.
% The aim of the diagnosis stage is to convey to the viewer, 
% on the one hand, which questions regarding the disease should be answered by the 
% diagnosis and which technical modalities are used for this purpose. 
%% ============================================================================
%% ============================================================================
Each diagnostic question is illustrated as an individual scene using the 
\textit{magazine style}. 
%% ============================================================================
Important questions for liver cancer include, e.g., the size and location of 
tumors as well as the the exact tumor type. 
%% ============================================================================
For this purpose, the 3D translucent liver model including the tumors is shown. 
%% ============================================================================
The tumors are enriched by \textit{glyphs} and \textit{annotations} to visually 
communicate the the main aspect of each related question (see 
Figure~\ref{fig:liver} (C2-C3)). 
%% ============================================================================
Simple and clean \textit{vector illustrations} describe diagnostic procedures, e.g., 
liver biopsy to determine the exact tumor type (see Figure~\ref{fig:liver} (D3)). 
%% ============================================================================
Procedures with anatomical views different than previously presented include a 
\textit{rotating 3D navigator model} of the organ that helps for view orientation.
%% ============================================================================

%% ============================================================================
The size and location of aneurysms in the brain vessels is also a critical 
diagnostic question, which we handle slightly differently (see 
Figure~\ref{fig:aneurysm} (C)). 
%% ============================================================================
This entails beginning with a localizing \textit{overall view} of the brain vessels 
with the translucent surrounding skull %in a translucent surrounding material 
before a \textit{zoom and rotation} transformation focuses the camera tightly on 
the aneurysm to emphasize its location and shape. 
%% ============================================================================
\textit{Panning and zooming} allows the user to get even closer to the aneurysm, 
a detail view is shown in the form of a 3D aneurysm model, with blood flow suggested by 
\textit{animated particles}, see Figure~\ref{fig:aneurysm}~(D). 
%% ============================================================================

%% ============================================================================
Additional scenes give an impression of diagnostic imaging modalities used. 
%% ============================================================================
Here, the \textit{magazine style} is again used to combine image and text 
information. 
%% ============================================================================
For each modality, a \textit{real image} is incorporated as an example, e.g., 
ultrasound, CT, and MRI in liver cancer, where in each the tumors are highlighted 
by contours, see Figure~\ref{fig:liver} (A4). 
%% ============================================================================
We use similar concepts to show imaging modalities employed in brain 
aneurysm and pelvic fracture diagnosis.  
%% ============================================================================

%% ============================================================================
\noindent 
\textbf{Treatment.} The treatment stage provides an overview of typical therapy 
options and key aspects that influence treatment decisions. 
%% ============================================================================
We do not consider rarely performed treatments that can only be done in special centers.
%relevant for our target audience.
%% ============================================================================

%% ============================================================================
The first treatment scene uses a \textit{flow chart} combined with 
\textit{long-form textual descriptions}. 
%% ============================================================================
The chart in form of a directed graph describes the basic treatment 
approaches and their key aspects. 
%% ============================================================================
By clicking on one of its nodes, the user gets more information in a 
\textit{magazine style} about the selected treatment. 
%% ============================================================================
Each treatment is shown as a \textit{2D vector illustration} 
of its key moment to provide clarity.
%% ============================================================================
We again use a \textit{navigator icon} with \textit{labels} to indicate key 
aspects of the procedure.
%% ============================================================================
An exception to this are the metal implants used in fracture treatment, such as 
in the pelvis data (see Figure~\ref{fig:pelvis} (D)). 
%% ============================================================================
Similar to bones in a CT scan, these can be easily visualized by DVR, where optimal 
views from different perspectives can be shown. % to the viewer. 
%% ============================================================================
In case of palliative treatments where the target is more towards symptom 
alleviation, we use \textit{icons} to create consistency and repetition between this 
stage and the earlier symptom stage. 
%% ============================================================================

%% ============================================================================
For example, in liver cancer there are essentially two treatment strategies, 
curative and palliative~\cite{Anwanwan2020} %which we show as a flow chart, 
(see Figure~\ref{fig:liver} (B4)). 
%% ============================================================================
To the right, we detail the key aspects that determine treatment in list form. This includes: the  number of tumors present, their size, whether they have grown into blood vessels or into surrounding tissue, whether they have already metastasized, how functional the liver still is, and the patient's general health. 
%% ============================================================================
On selection of the curative therapy chart element, the user is directed 
through the set of curative therapies. 
%% ===========================================================================
Here, the user learns that the goal of curative therapy is to cure the cancer, 
methods of which include (1) surgical removal of the tumor(s), (2) ablation, 
in which the tumor(s) is destroyed by heat or microwaves, and (3) radiation 
(Figure~\ref{fig:liver} (C4-A5)). 
%% ===========================================================================
The user then is taken through the palliative treatment scenes. %, where the cancer 
%has progressed too far and a cure is no longer possible. 
%% ===========================================================================
We use %a sequence of 
icons to indicate alleviation of symptoms that were 
introduced earlier in the story, as well as new icons, such as those for 
chemotherapy, cytostatic drugs, and high calorie foods that may prolong the 
patient's life and relieve symptoms (see Figure~\ref{fig:liver} (B5)). 
%% ============================================================================

%% ============================================================================
\noindent
\textbf{Prognosis.} In addition to general statistics, the prognosis of a disease 
also includes several parameters depending on the severity and the chosen 
treatment. 
%% ============================================================================
Since a detailed presentation of all dependencies and resulting parameters would 
overwhelm the general user, we limit ourselves to the most important 
parameters to give insight into the liklihood of a cure. 
%% ============================================================================
Based on the \textit{partitioned poster} genre, we produce an 
\textit{infographic-style illustration} that makes use of \textit{color} and 
\textit{laddered text} to aid in information absorption and memory retention of 
key prognostic information. 
%% ============================================================================
% We use \textit{glyphs} for men and women to rapidly identify their personal 
% prognostic information. 
We use \textit{glyphs} for men and women to convey that, unlike incidence, there 
are no significant prognostic differences between men and women. 
% Based on the \textit{partitioned poster} genre, most of these facts are represented 
% by \textit{annotated information graphics} and \textit{symbols}. 
% %% ============================================================================
% These visuals give a clear, succinct overview of statistical parameters without 
% going in too much detail. 
%% ============================================================================

%% ============================================================================
The prognosis for liver cancer depends on the cancer stage and the condition of 
the liver~\cite{Anwanwan2020}. 
%% ============================================================================
We emphasize the relative 5-year survival, which is around 18\,\% for men and 
women, in the \textit{largest typeface} with an \textit{attention-drawing accent color} 
along with \textit{icons} indicating men and women (see Figure~\ref{fig:liver} (C5)). 
%% ============================================================================
We use a \textit{smiley symbol} to indicate the group where tumor removal often 
experiences a positive outcome. 
%% ============================================================================
For stage I tumors, %(single tumors, without blood vessel or lymph node involvement 
%and without metastases), 
the 5-year relative survival is around 62\,\% 
in women and around 54\,\% in men. 
%% ============================================================================
In stage IV, %(lymph node involvement or metastases), 
however, it is only 2\,\%. 
%% ============================================================================
We repeat the use of larger typeface with accent color for the survival with 
gender symbols at a slightly smaller size. %in these two groupings. 
%% ============================================================================
For both the aneurysm and pelvic fracture stories we use a similar presentation of 
laddered text with accent colors for percentages with symbols to indicate 
affected genders. 
%% ============================================================================

%% ============================================================================
\noindent
\textbf{Prevention.} In this stage, we focus on illustrating avoidable risk factors 
to give the user a sense of agency. 
%% ============================================================================
Risk factors such as age or genetic factors that a person cannot influence 
are excluded since they are not actionable for the user. 
%% ============================================================================
Similar to the symptoms, we use the \textit{partitioned poster} genre and 
utilize \textit{icons} to better recognize and understand the presented 
information, e.g., a martini glass for the recommendation to reduce 
alcohol consumption.
%% ============================================================================
% For each risk factor, we artistically create a \textit{symbols} combined with 
% a \textit{caption}, which are displayed one after the other. 
%% ============================================================================

%% ============================================================================
%Visiting our running example for a final time, 
The main risk factor for liver cancer is cirrhosis, caused typically by chronic 
hepatitis B virus infection, depicted with a \textit{syringe} and \textit{caption} 
to vaccinate against hepatitis B, or high alcohol consumption, which we depict with an 
\textit{alcoholic beverage icon} and a \textit{caption} to limit alcohol 
consumption (Figure~\ref{fig:liver} (D5)). 
%% ============================================================================
Another risk factor is obesity which is depicted by a \textit{scale symbol} with 
an \textit{arrow} indicating increasing weight along with the caption 
``Keep weight in a healthy range.''. 
%% ============================================================================
% Non-alcohol-related fatty liver disease, which also increases the risk of liver 
% cancer, becomes increasingly important. 
% %% ============================================================================
% It can be a consequence of diabetes mellitus, often triggered by obesity. 
% %% ============================================================================
% We use a \textit{scale symbol} with an \textit{arrow} indicating increasing weight along with 
% the caption ``Keep weight in a healthy range.'' 
%% ============================================================================
Smoking also increases the risk of the disease, which we depict with a \textit{barred 
cigarette icon} with the \textit{caption} ``Quit smoking.''
%% ============================================================================
We can similarly use such icons to show risk factors for the brain aneurysm 
and pelvic fracture stories.

\section{Discussion}
\label{sec:discussion}
%% ============================================================================
%% ============================================================================
In the following, we outline general issues that arise during medical story creation. 
%% ============================================================================
We cover principle decisions regarding scene design, such as story element style
and underlying design patterns.
%as well as their validation. 
%summarize basic genre and design patterns and discuss possible alternatives. 
%% ============================================================================
Furthermore, we discuss necessary story adaptations to address other diseases or 
other communication goals, e.g., education. 
%% ============================================================================

%% ============================================================================
\noindent\textbf{Content production.} For our purposes of showing an initial 
concept for narrative medical visualization, the functionality of PowerPoint for story creation was sufficient. 
%% ============================================================================
However, its content production feature set remains limited, with significant manual effort required to integrate multiple media types to tell a comprehensive story. We used several external tools to bridge this gap, including Adobe Illustrator to produce icons and treatment illustrations. While PowerPoint is able to embed 3D surface models, medical image feature extraction and advanced visualization, e.g., DVR, are not supported, which led to a need to use additional software.
%% ============================================================================
%However, this involved significant manual effort. PowerPoint does not provide a way to prepare medical image data and extract structures. Therefore, several other tools had to be used to segment and simulate the data. Furthermore, tools like Adobe Illustrator had to be used to create the 2D illustrations and icons. Integrating multiple media types to tell a comprehensive story was therefore very difficult in PowerPoint. Besides, techniques such as DVR are not supported in PowerPoint. Therefore, we had to import predefined videos and images to show the pelvis structures, for example.
%% ============================================================================
% In the future, efficient tools that provide extensive functionalities in terms 
% of data preprocessing, visualization and animation generation will be essential 
% to drive the application of narrative techniques to medical data. 
%% ============================================================================
 
%% ============================================================================
\noindent\textbf{Scene design.} Following the suggestion of 
B{\"o}ttinger et al.~\cite{bottinger2020challenges} to make scenes "as simple 
as possible without risking scientific credibility," we extract as much information 
as possible from real medical image data to generate data-driven stories. 
%% ============================================================================ 
However, for creating simple and clearly understandable scenes for our target
audience, visual abstraction and easy-to-use interaction techniques are necessary~\cite{Viola2020visual}. 
%% ============================================================================
We used textual and verbal descriptions and avoided technical terms where possible. 
%% ============================================================================
We added 2D vector illustrations to show important treatment concepts 
in a simplified way. 
%% ============================================================================ 
While 3D data-driven models with surgical instruments would also be possible, this 
creates visually complex scenes that we felt may overwhelm or scare the user. 
%% ============================================================================

%% ============================================================================
For touch-based interaction, we use interaction types that do not require a high degree of accuracy. 
%% ============================================================================
These include single-touch gestures, e.g., rotation around a predefined axis 
and clicking on objects, as well as familiar multi-touch gestures, such as 
panning and zooming. %that most users are familiar with from using everyday objects such as smartphones and laptops. 
%% ============================================================================
Such interactions take advantage of user familiarity with everyday objects 
such as smartphones and tablet PCs without requiring extra equipment, such as a 
mouse or keyboard.
%% ============================================================================
%None of these interactions require high accuracy; the advantages of direct natural interaction outweigh the disadvantages. 
%% ============================================================================
%In addition, there is the smaller space requirement compared to, for example, a mouse-based interaction.
%% ============================================================================

To select an appropriate narrative genre, there are currently no guidelines on which genre is most 
suitable for a given context. 
%% ============================================================================
Since, we want an interactive and multi-media environment for our stories, 
slideshows seemed most appropriate.
%% ============================================================================
% From our experience slideshows are used, e.g., on an automated loop at trade shows. 
% %at trade shows without presenter.
% %% ============================================================================
% Thus, this genre is suitable for asynchronous stories when more complex contexts and processes are described. 
% %% ============================================================================
% This fits our goal of informing the general public (asynchronous story), as the communication of disease data consists of several related aspects that create a comprehensive overall picture. 
%% ============================================================================
Within the scenes, we often combine 2D/3D representations with textual descriptions 
arranged in the magazine style to explain disease stages in a short and memorable way. 
%% ============================================================================
In principle, we could also use another genre for the story or parts of the story, 
such as comics. 
%% ============================================================================
%However, comics are not really meant for interactive, multi-media stories. 
%% ============================================================================
Thus, a validation and derivation of guidelines of contextual genre 
recommendations is an important point for future research.  

%% ============================================================================
% This way, depicted objects are explained in a short and memorable way. 
% %% ============================================================================
% At other points, such as the depiction of symptoms and risk factors, no textual 
% explanations are necessary. 
% %% ============================================================================
% Memorable icons combined with annotations, arranged in a poster style, should 
% convey the most important facts in an understandable way. 
% %% ============================================================================
% At this point, other genres such as comics would have been possible. 
%% ============================================================================

%% ============================================================================
To inform the general public about a disease, we use all design patterns defined 
by Bach et al.~\cite{Bach2018} (cf. Section~\ref{subsec:narpatterns}). 
%% ============================================================================
The two most important patterns in our stories are \textit{framing by hiding data} 
and \textit{structuring by revealing data}. 
%% ============================================================================
For example, within the 3D models only the most important structures are depicted 
while structures irrelevant to the story are excluded, e.g., other abdominal organs 
in the liver scenario. % or the brain in the aneurysm story. %, and redacted unnecessary information from the data-driven 3D models, e.g., the liver cancer dataset contained cysts, which we removed for this story. 
%% ============================================================================
%Structures, such as other abdominal organs in the liver scenario or the brain for the aneurysm story, which are not essential for understanding, are not shown. 
%% ============================================================================
In addition, combinations of 3D structures, annotations/glyphs, and textual 
descriptions are revealed sequentially to avoid confronting the user with 
a visual overload. 
% 
%% ============================================================================
We use emotional and engaging patterns to capture and maintain the user's attention. 
%% ============================================================================
We use an initial patient description to help users relate to the story, and incorporate interactive components encouraging users to take action. 
%% ============================================================================
Argumentative patterns in the form of memorable icons convey avoidable risk factors. 
%% ============================================================================
Our aim is to raise awareness of a disease and to encourage taking action and adopting a healthier lifestyle. 
%% ============================================================================

\noindent\textbf{Evaluation of Medical Stories.} Our stories were designed by 
scientists with many years of experience in the visualization of medical data. 
%% ============================================================================
One of our co-authors is a medical illustrator and thus brings a lot of experience 
regarding the design of illustrations for the general public. 
%% ============================================================================
In this forward-looking paper, we have basically conceptualized what 
medical stories could look like. 
%% ============================================================================
We have not yet done any evaluation with the intended audience which is fundamental for future work. 
%% ============================================================================

%% ============================================================================
\noindent\textbf{Communication goals.} In addition to informing, educating an 
audience would be another important communication goal. 
%% ============================================================================
For example, one possible scenario would be to teach students about various medical conditions. 
%% ============================================================================
The use of engaging patterns in the form of interactive exploration to 
maintain students' enjoyment during the learning process is particularly 
important~\cite{Rheingans2020}. 
%% ============================================================================
An option could be to integrate an interactive quiz, where the user could check 
how much they have learned from the story. 
%% ============================================================================
Moreover, argumentative patterns such as spaced repetition should be used to help the 
user remember the most important facts. 
%% ============================================================================

%% ============================================================================
Furthermore, details on demand would be useful for learning. 
%% ============================================================================
In currently available apps, such as "language learning apps", the respective topic is explained using a meaningful example.
%% ============================================================================
In addition, the user has the option of viewing further language-specific 
examples to consolidate the learning material. 
%% ============================================================================
This details-on-demand principle could be applied to medical data. 
%% ============================================================================
Upon request, the user could see additional records, e.g. a stage IV liver cancer 
dataset, with other anatomical features. 
%% ============================================================================

%% ============================================================================
\noindent\textbf{Application to other diseases.} In this paper, we focus on 
diseases which can be diagnosed based on radiological image data. 
%% ============================================================================
However, several diseases require other technologies for diagnosis. 
%% ============================================================================
Examples are respiratory diseases such as asthma, heart rhythm disorders, or 
infectious diseases such as HIV, which are diagnosed by pulmonary function tests, 
electrocardiogram, and blood tests, respectively. 
%% ============================================================================
While our basic division into the seven stages can also be applied to such diseases, 
the media used must be adapted accordingly.  
%% ============================================================================
For example, 3D models based on real data cannot be generated for infectious 
diseases. 
%% ============================================================================
Similarly, treatment options would have to be adapted. The division into curative 
and palliative methods is not always possible, since many diseases cannot be 
cured. % at present.  
\section{Research Agenda}
\label{sec:agenda}
%% ============================================================================
%% ============================================================================
This section focuses on open research challenges focusing on the combination of 
narrative techniques and medical data. 
%% ============================================================================
% While storytelling has already been used for other scientific data to bring 
% research results closer to the general public, it has hardly been used for medical 
% data. 
%% ============================================================================
%% ============================================================================

% However, regarding the design of visualizations for specific audiences and having
% simplicity in mind, we may have to think even more deeply about the colormap we are
% using. For audience-oriented colormap design, we have to choose the type of the colormap
% (continuous vs. discontinuous), about the number of hues or colors, and about
% psychological effects of specific colors.

\noindent
\textbf{Authoring tools for narrative medical visualization.}
%% ============================================================================
When presenting medical information such as incidence rates or prognostic factors, any authoring tool suitable for data story creation would be suitable. 
However, presentation of medical imaging data offers additional challenges.
Providing efficient authoring tools tailored to medical imaging data is a key aspect of 
advancing the use of narrative techniques in this area. 
%% ============================================================================
%% ============================================================================
%% ============================================================================
Such an authoring tool would need to support data preprocessing (e.g.,
segmentation, smoothing, filtering) and provide techniques to add narrative 
elements. % to the processed data. 
%% ============================================================================
Here, also advanced labeling~\cite{Oeltze2014} and animation creation~\cite{Preim2020} 
techniques for medical data are needed. 
%% ============================================================================
Existing expert tools for analyzing medical data, e.g., MeVisLab~\cite{MeVisLab}, 
3D slicer~\cite{Kikinis2014}, or VTK~\cite{Schroeder1998} 
could be chosen as a basis. % for this. 
%% ============================================================================
These are powerful medical image processing and visualization frameworks freely
available for non-commercial research.
%% ============================================================================
Their modular character allow fast integration and testing of new algorithms. 
%% ============================================================================
They could be extended with narrative modules which allow scene generation scenes based on a selected scene type and transitions between scenes. 
%% ============================================================================
Key features should be interaction tracking, e.g., to create animations 
and the definition of which user interactions should be provided. % with the data. 
%% ============================================================================

%% ============================================================================
An alternative would be to provide an authoring tool to create web-based stories. 
%% ============================================================================
This would allow even more people to access the stories.
%for information or education. 
%% ============================================================================
For the creation of web-based stories, libraries such as VMTK~\cite{Izzo2018} or VTK.js could be combined for the preparation and visualization of imaging data and D3~\cite{d3js} for the incorporation of information visualization and interaction. 

\noindent
\textbf{General pattern design for medical data.} The definition of 
dedicated patterns to capture medical narratives helps authors to structure their stories. 
%% ============================================================================
In this paper, we propose a pattern for the narrative presentation of 
disease data comprising seven stages. 
%% ============================================================================
However, there are many other medical aspects, such as healthy metabolic processes, 
pregnancy, and medical procedures, for which this pattern would not be suitable. 
%% ============================================================================
Other patterns should be derived to structure such topics.
%the range of medical scenarios has to be grouped to classes.
%% ============================================================================
%For each class a general pattern should be derived to structure the stories. 
%% ============================================================================

%% ============================================================================
\noindent
\textbf{Narrative medical visualization for patients.} 
%% ============================================================================
In addition to medically interested people, patients could benefit from narrative medical visualization. 
%% ============================================================================
Patients come into contact with medical data through their physicians who explain planned procedures and diagnosed diseases, including 
possible treatments and their prognosis. 
%%% ============================================================================
The patient typically sees excerpts from imaging data or schematic drawings on information sheets. % of planned diagnostic 
%and therapeutic procedures. 
%% ============================================================================
However, since patients often have little medical knowledge, communication difficulties can occur, which could lead to uncertainty and fear. % on the part of the patient. 
In a doctor-patient conversation, the patient could instead be informed about his or her 
condition using personalized visual representations. 
%% ============================================================================
Here, narrative visualization could help the patient to understand diagnostic 
and therapeutic procedures such that a suitable therapy plan can be made in 
consultation with the physician. This would strengthen opportunities for participatory medicine.

\noindent
\textbf{Support experts with narrative medical visualization.} 
%% ============================================================================
% At the beginning, we defined different user groups that come into contact with 
% medical data. 
%% ============================================================================
In addition to the general public, medical experts are also a potential target group for 
narrative visualization. 
%% ============================================================================
Many expert visualization tools partly consist of very complex visual representations where the experts 
need extensive explanations and there is a steep learning curve. 
%% ============================================================================
Especially for prospective experts, such as medical students, narrative techniques 
could help to understand complex medical data~\cite{Rheingans2020}. 
%% ============================================================================
In contrast to stories developed for a general audience, expert narratives would need more emphasis on data preservation and, consequently, less abstraction. 
%% ============================================================================
It would be interesting to further explore how expert tools can be enriched with narrative techniques 
and what impact this has on understanding. % the data.
%% ============================================================================

%Story Telling for Student Education and Expert tools such as Sectra table

% \noindent
% \textbf{Memorable narrative medical visualization.}
% Memorability for visualization: What are the key elements for making a memorable visualization? This is still an immature research direction.~\cite{Tong2018,Kosara2013}

\noindent
\textbf{Narrative medical visualization evaluation.} 
%% ============================================================================
To assess the usefulness of storytelling in a medical context, it is important to measure the quality 
of the visualization w.r.t. achieving communication goals. 
%% ============================================================================
Evaluation of narrative visualizations is a difficult and complex task with 
which only a few works have currently dealt~\cite{Mahyar2015,Boy2015}. 
%% ============================================================================
The basic idea is to evaluate \textit{design decisions made}%(e.g., the rendering 
%styles or color scales chosen and the level of interaction allowed)
, \textit{techniques employed}%(e.g., animations, textual descriptions, audio)
, and \textit{media used} %(e.g., static images, interactive displays, and virtual environments) 
in terms of aspects such as comprehensibility~\cite{Figueiras2014tell}, 
memorability~\cite{Borkin2015}, and effectiveness in engaging a general audience~\cite{Boy2015,Mckenna2017}. 
%% ============================================================================

%% ============================================================================
In contrast to medical expert visualizations, which are often validated in controlled 
lab studies, typically within a short time frame, evaluation 
of narrative medical visualization will require a different approach to 
account for various scenarios and to reflect real-world uses.
%% ============================================================================
To get meaningful results, we have to reach a wider and more diverse 
audience than the usual small group of experts used in lab studies. 
%% ============================================================================
Kosara et al.~\cite{Kosara2013} suggested to use crowdsourcing platforms 
%such as Amazon’s Mechanical Turk, 
for such larger visualization studies. 
%% ============================================================================
In addition, objective measures of comprehensibility, memorability, and user-engagement 
have to be defined. 
%% ============================================================================
Eye-tracking studies could help to better understand how different medical-related 
audiences view and interact with visualizations. 

\noindent
\textbf{Advanced media for narrative medical visualization.} 
%% ============================================================================
Many of the existing visualization and interaction approaches were originally 
designed for a standard desktop computer set-up with mouse and keyboard input. 
%% ============================================================================
However, in the last years new advances have been made in human-computer 
interaction technology~\cite{Besanccon2021}. 
%% ============================================================================
% New display technologies such as spherical displays are well suited to visualize 
% and interact with geo-scientific data~\cite{Bottinger2020}. 
%% ============================================================================
New display technologies such as multi-touch displays have become very popular 
as they are intuitive, easy to use, and allow a direct manipulation of the visual 
data representation by multiple users. 
%% ============================================================================
Furthermore, VR environments are increasingly being developed for medical 
scenarios such as surgical planning and medical education. 
%% ============================================================================
It would be interesting to investigate how these mediums could be combined 
with narrative techniques to communicate medical data to general audiences. % and 
%what advantages and disadvantages it would have compared to a conventional touch 
%display. 
%% ============================================================================
Furthermore, concepts such as voice and gesture control as well as human avatars 
to answer questions may bring complex scientific data 
closer to a general audience~\cite{Ynnerman2020}. 
%% ============================================================================
A more detailed investigation is needed regarding their suitability for medical 
data.

\section{Conclusion} 
\label{sec:conclusion}
%% ============================================================================
%% ============================================================================
The use of narrative techniques represents an important research focus with the 
goal of communicating complex scientific data in an understandable way to a general 
audience without specific expertise. 
%% ============================================================================
To date, however, there are few approaches leveraging narrative techniques for 
medical data, although the audience for this is large. 
%% ============================================================================
Patients, relatives and people interested in medicine can benefit from custom
medical representations and expand their knowledge. 
%% ============================================================================
In addition, physicians and medical students could benefit 
from the use of narrative techniques. 
%% ============================================================================

%% ============================================================================
In this paper, we provide a first proof of concept demonstrating how narrative techniques could help explain disease characteristics to a wider audience. 
%% ============================================================================
The combination of data-driven and illustrative presentations seems to be a 
powerful way to present complex diagnosis and treatment methods in an 
understandable way. 
%% ============================================================================
Important points for future research are the development of appropriate 
authoring tools and the validation of narrative medical presentations.

% \section*{Acknowledgments}
% This work was partially funded by the Carl Zeiss Foundation and the Federal Ministry for Economic Affairs and Energy of Germany. The authors like to thank Philipp Berg and Samuel Vo{\ss} for the fruitful discussions on these and related topics.

%-------------------------------------------------------------------------
% bibtex
%\bibliographystyle{eg-alpha-doi} 
% \bibliography{egbibsample}       

% biblatex with biber
\printbibliography[keyword=primary]
%-------------------------------------------------------------------------

%Append keywords to identify different bibliography entries.
% appendstrict only appends if the field is nonempty,
% we use that to add a comma to avoid mushing together two keywords
% \DeclareSourcemap{
%   \maps[datatype=bibtex, overwrite]{
%     \map{
%       \perdatasource{biblatextest1.bib}
%       \step[fieldset=KEYWORDS, fieldvalue={, }, appendstrict]
%       \step[fieldset=KEYWORDS, fieldvalue=primary, append]
%     }
%     \map{
%       \perdatasource{biblatextest2.bib}
%       \step[fieldset=KEYWORDS, fieldvalue={, }, appendstrict]
%       \step[fieldset=KEYWORDS, fieldvalue=secondary, append]
%     }
%   }
% }

% \DeclareFieldFormat{labelnumberwidth}{\mkbibbrackets{#1}}
% \renewbibmacro*{cite}{%
%   \printtext[bibhyperref]{%
%     \printfield{labelprefix}%
%     \ifkeyword{secondary}
%       {\printfield{labelnumber}}
%       {\printfield{labelalpha}%
%       \printfield{extraalpha}}}}

% \defbibenvironment{bibliographyNUM}
%   {\list
%      {\printtext[labelnumberwidth]{%
%         \printfield{labelprefix}%
%         \printfield{labelnumber}}}
%      {\setlength{\labelwidth}{\labelnumberwidth}%
%       \setlength{\leftmargin}{\labelwidth}%
%       \setlength{\labelsep}{\biblabelsep}%
%       \addtolength{\leftmargin}{\labelsep}%
%       \setlength{\itemsep}{\bibitemsep}%
%       \setlength{\parsep}{\bibparsep}}%
%       \renewcommand*{\makelabel}[1]{\hss##1}}
%   {\endlist}
%   {\item}

 \newrefcontext[sorting=none]
 \printbibliography[env=bibliographyNUM, title={Additional References}, keyword=secondary, resetnumbers]
%\printbibliography[title={Additional References}, keyword=secondary]

\end{document}